\documentclass[10pt,sigconf,letterpaper]{acmart}
\usepackage{graphicx}
\usepackage{subfigure}
\usepackage{lipsum}
\usepackage{multirow}  
\usepackage{tikz}
\usepackage{breakurl}
\usepackage{amsmath}
\usepackage[english]{babel}

\setcopyright{none}
\copyrightyear{}
\acmYear{}
\acmDOI{}
\acmPrice{}
\acmISBN{}
\renewcommand\footnotetextcopyrightpermission[1]{}

\usepackage{csquotes}

\usepackage[nice]{nicefrac} 
\usepackage{xcolor}         
\usepackage{xifthen}
\usepackage{mathtools}
\usepackage{multicol}
\usepackage{placeins}

\usepackage[utf8]{inputenc}
\usepackage{hyperref}
\usepackage{color}
\usepackage{subfigure}
\usepackage{graphicx}
\usepackage{multirow}
\usepackage{pifont}

\usepackage{soul}
\usepackage[inline,shortlabels]{enumitem}
\usepackage{threeparttable}
\usepackage{colortbl}
\usepackage{xcolor}
\usepackage{enumitem}
\setlist[itemize]{align=parleft,left=0pt..1em}

\newif\ifremark
\long\def\remark#1{
\ifremark%
	\begingroup%
	\dimen0=\columnwidth
	\advance\dimen0 by -0.25in%
	\setbox0=\hbox{\parbox[b]{\dimen0}{\protect\em #1}}
	\dimen1=\ht0\advance\dimen1 by 2pt%
	\dimen2=\dp0\advance\dimen2 by 2pt%
	\vskip 0.25pt%
	\hbox to \columnwidth{%
		\vrule height\dimen1 width 3pt depth\dimen2%
		\hss\copy0\hss%
		\vrule height\dimen1 width 3pt depth\dimen2%
	}%
	\endgroup%
\fi}
\usepackage{pgf-pie}
 
\newcommand*\halfcirc[1][1ex]{%
  \begin{tikzpicture}
  \draw[fill] (0,0)-- (90:#1) arc (90:180:#1) -- cycle ;
  \draw (0,0) circle (#1);
  \end{tikzpicture}}
\newcommand*\fullcirc[1][1ex]{\tikz\fill (0,0) circle (#1);} 
\remarktrue

\begin{document}

\title[A Fresh Look at ECN Traversal in the Wild]{A Fresh Look at ECN Traversal in the Wild}

\author{Hyoyoung Lim}
\email{hyoyoung.lim@colorado.edu}
\affiliation{%
  \institution{University of Colorado Boulder}
}

\author{Seonwoo Kim}
\email{seonwoo.kim@colorado.edu}
\affiliation{%
  \institution{University of Colorado Boulder}
}

\author{Jackson Sippe}
\email{jackson.sippe@colorado.edu}
\affiliation{%
  \institution{University of Colorado Boulder}
}

\author{Junseon Kim}
\email{jskim@unist.ac.kr}
\affiliation{%
  \institution{UNIST}
}

\author{Greg White}
\email{g.white@cablelabs.com}
\affiliation{%
  \institution{CableLabs}
}

\author{Chul-Ho Lee}
\email{chulho.lee@txstate.edu}
\affiliation{%
  \institution{Texas State University}
}

\author{Eric Wustrow}
\email{ewust@colorado.edu}
\affiliation{%
  \institution{University of Colorado Boulder}
}

\author{Kyunghan Lee}
\email{kyunghanlee@snu.ac.kr}
\affiliation{%
  \institution{Seoul National University}
}

\author{Dirk Grunwald}
\email{dirk.grunwald@colorado.edu}
\affiliation{%
  \institution{University of Colorado Boulder}
}

\author{Sangtae Ha}
\email{sangtae.ha@colorado.edu}
\affiliation{%
  \institution{University of Colorado Boulder}
}

\begin{abstract} 
The Explicit Congestion Notification (ECN) field has taken on new importance due to Low Latency, Low Loss, and Scalable throughput (L4S) technology designed for extremely latency-sensitive applications (such as cloud games and cloud-rendered VR/AR). ECN and L4S need to be supported by the client and server but also all devices in the network path. We have identified that “ECN bleaching”, where an intermediate network device clears or “bleaches” the ECN flags, occurs and quantified how often that happens, why it happens and identified where in the network it happens.

In this research, we conduct a comprehensive measurement study on end-to-end traversal of the ECN field using probes deployed on the Internet across different varied clients and servers. Using these probes, we identify and locate instances of ECN bleaching on various network paths on the Internet. In our six months of measurements, conducted in late 2021 and early 2022, 
we found the prevalence varied considerably from network to network.
One cloud provider and two cellular providers bleach the ECN field as a matter of policy.
Of the rest, we found 1,112 out of 129,252 routers, $4.17\%$ of paths we measured showed ECN bleaching.
\end{abstract}

\maketitle

\section{Introduction}\label{sec:intro}

RFC 3168~\cite{ramak} specifies the incorporation of Explicit Congestion Notification (ECN) to TCP/IP, which allows routers to signal 
impending congestion to endpoints before buffers overflow or packets get dropped by routers. 
The IP header contains a 2-bit field known as the ECN field to enable such an end-to-end notification of impending network congestion without dropping packets.
To enable ECN, a sender marks its packets as ECN-Capable and then reacts to congestion signaled by the network.  
As packets traverse the network, congested devices mark ECN-enabled packets
to indicate congestion rather than dropping those packets.
The receiver monitors received packets for ECN congestion signals and sends congestion feedback to the sender.
ECN relies on the network devices to either pass along packets with the ECN information unmodified if there is no congestion or to indicate congestion by marking the ECN bits.
It only takes one misbehaving device in a network path to eliminate the benefit of ECN -- 
if any network device between a sender and receiver clear the ECN bits, the sender and receiver
will not learn about the impending congestion.  

There are several benefits to ECN~\cite{rfc8087}, particularly as a congestion signal 
that reduces application latency by foregoing dropping packets to signal congestion. 
Congestion control algorithms treat ECN congestion signals identically to packet drops as far as congestion response.
Most OSs have ECN enabled by default~\cite{learmonth2016pathspider,mandalari2018measuring}, meaning they respond to and use ECN if requested by the sender.

Recently, the L4S (Low Latency, Low Loss, and Scalable throughput) standard proposal~\cite{ietf-tsvwg-l4s-arch-17}
recognized that the original definition of ECN could be improved
to significantly reduce latency and latency variation, and could improve the scalability of congestion control designs for wide-area networks using
the mechanisms previously used in Datacenter TCP~\cite{10.1145/1851182.1851192,DBLP:journals/rfc/rfc8257}
while coexisting with the existing TCP and QUIC traffic on the Internet.
L4S aims to enable high-bandwidth, extremely latency-sensitive applications, such as cloud games and cloud-rendered VR/AR, to achieve their performance objectives to a degree that has not been practically feasible with existing congestion feedback mechanisms
and is a key element of proposed access network technologies such as Low-Latency DOCSIS.
L4S uses the ECT(1) codepoint as an identifier for `L4S' packets to be distinguished from `classic ECN' packets and has a separate treatment of L4S packets via an independent queue to reduce queuing delay~\cite{ecn2020Schepper}.\footnote{The ECT(1) codepoint was originally defined along with ECT(0) to indicate ECN-Capable Transport (ECT) in the ECN specification~\cite{ramak}.}
Unfortunately, some network devices clear or ``bleach'' the ECN bits, limiting the benefit
of ECN and also L4S.

ECN is widely available since most server OSs have ECN enabled by default~\cite{learmonth2016pathspider,mandalari2018measuring}.
Studies~\cite{bauer2011measuring,trammell2015observing,mandalari2018measuring}
using path traversal of ECN using a \texttt{traceroute}-based method have shown
that path traversal of ECN has become close to universal (as long as two endpoints have ECN enabled), but there are still a non-trivial fraction of paths along which network nodes wipe the ECN field of packets. 
In 2019, Roddav \textit{et al.}~\cite{roddav2019usage} used passive measurements to show that only 5.23\% of traffic is using ECN codepoints, possibly because of ECN ``bleaching''.

It remains largely unknown where such bleaching paths are prevalent in the Internet in terms of ISPs and geographical regions, and why they exist or if ECN signaling is being used.
Mandalari \textit{et al.}~\cite{mandalari2018measuring} observed that more than half of the mobile carriers that they tested bleach the ECN field to zeros, and the ECN bleaching cases happen in the first hop from mobile clients.
Two recent studies ~\cite{Padma2027apple,Holland2020} indicate that there are a few networks that have enabled ECN congestion signaling, but the majority have not.

In light of significant interest in the new L4S ECN architecture, it is now the time to revisit the usability (i.e. traversal) of ECN in cellular as well as wired networks.
We conducted a study of a 24 hour passive trace of network traffic at a single vantage point within a University network, looking at the prevalence of the different values within ECN fields in the IP and TCP packet headers. 
We also have conducted an active measurement study on end-to-end traversal of the ECN field using probes deployed on the Internet across different locations and providers using two forms of active measurements.
First, we used PATHspider~\cite{learmonth2016pathspider} to determine
which HTTP/HTTPS servers have ECN enabled to better understand ECN adoption.
Then, using a tool we developed, we identified and located instances of ECN bleaching on various network paths on the Internet.

We learned different things from our passive and active measurements.
In our six months of measurements, conducted in late 2021 and early 2022, aside from one cloud provider that bleaches the ECN field as a matter of policy, we found 1,112 out of 129,252 routers, and 4.17\% of paths showed ECN bleaching and many networks have no bleaching.
However, previous studies~\cite{roddav2019usage} using passive measurements
showed that only 5.23\% of traffic is using ECN, which is a perplexing
result given that ECN is enabled for most modern OSs.
We found that although ECN is enabled by receivers for \emph{incoming} packets,
ECN is not requested for \emph{outgoing} packets by senders by default.
In other words, many networks and servers are ready for ECN but client
devices~\cite{enable-ecn} or applications~\cite{windows-ecn} need to enable ECN for it to be used.

\begin{table}[t]
\label{tbl:intro}
\caption{Dataset types and descriptions.}
\begin{center}
\scalebox{0.95}{%
\begin{tabular}{| c | l | r | r |}
\hline
Type & Data collection & \# of paths & Data size \\
\hline\hline
Active & Server measurement & 2,656,935 & 3.4 GB \\
\hline
Active & Traceroute localization & 530,795 & 2.4 GB \\
\hline
Passive & 1-day traffic trace data & - & 9.8 GB \\
\hline
\end{tabular}}
\end{center}
\end{table}

The remainder of this paper is structured as follows. First, we describe the old (classic) and new (L4S) schemes of ECN in Section 2. Then, in Section 3, we present the probes we use and the data collection methodology. Next, using the results from our methodology, we characterize the current state of ECN traversal and compare it to previous studies in Section 4. We discuss additional results in Section 5. Finally, Section 6 reviews related work before Section 7 concludes with a summary of our findings.

\section{Background}\label{sec:background}
\subsection{Explicit Congestion Notification}\label{subsec:ecn-background}

The original ECN specification (i.e. Classic ECN) was standardized in 2001 to allow routers to provide end-to-end notification of incipient network congestion instead of using packet drops as an indication of congestion~\cite{ramak}. The basic idea behind ECN is to provide an unambiguous signal of congestion without additionally causing a degradation (i.e. packet loss) in the transfer of data.  In a router, ECN is implemented via an active queue management (AQM) mechanism that evaluates queue depth and/or queue latency, and when the queue exceeds a threshold it marks the packet to indicate congestion has been experienced. The benefits of ECN include reducing packet loss in the Internet, which in turn leads to avoiding an increased application-layer latency caused by packet retransmissions. As a result, ECN improves the user experience in latency sensitive applications.

\begin{table}[t]
\caption{ECN codepoints and their meanings.}\label{bg:l4s_ect}
\vspace{-0.15in}
\begin{center}
\scalebox{0.95}{%
\begin{tabular}{| c | c | c | c |}
\hline
ECN codepoint  & Binary & L4S meaning \\
\hline\hline
Not-ECT        & 00 & Not ECN-capable transport\\
\hline
ECT(0)         & 10 & Classic ECN-capable transport\\
\hline
ECT(1)         & 01 & L4S-capable transport\\
\hline
CE             & 11 & Congestion experienced\\
\hline
\end{tabular}}
\end{center}
\end{table}

We provide a brief overview of how ECN works with 
TCP (on top of IP), assuming that two endpoints and all the 
intermediate routers between the endpoints are ECN-capable. 
ECN uses an ECN field of two bits in the IP header, which are 
the last two bits of the type of service (TOS) field originally 
defined in the IP header. The codepoints of the ECN field and 
their meanings are shown in Table \ref{bg:l4s_ect}. The ECT codepoints `10' 
and `01' -- ECT(0) and ECT(1) -- were originally equivalent in 
the ECN specification \cite{ramak}, but ECT(1) 
has been repurposed to indicate L4S packets under the L4S service 
architecture \cite{ietf-tsvwg-l4s-arch-17}, leaving ECT(0) to 
indicate Classic ECN packets. ECN also requires support from 
the transport protocol (i.e., TCP) in addition to the functionality 
provided by the ECN field of IP packets.

The endpoints, i.e., TCP client and server, first negotiate ECN capability during the establishment of their connection. After successful negotiation, a Classic ECN sender sends IP packets with ECT(0) set in the ECN field to indicate that they are ECN-capable. If a router detects impending congestion (with the help of an AQM mechanism) when a Classic ECN-capable packet arrives at the router, it marks the ECN field of the packet with the CE codepoint, and forwards the packet instead of dropping it. When the receiver receives this packet with CE set in the ECN field, it informs the sender of the congestion indication by sending its next TCP ACK with the ECN-Echo (ECE) flag marked in the TCP header. Upon receiving this TCP ACK, the sender then reacts to the congestion as if a packet has been dropped and sends the next packet with the Congestion Window Reduced (CWR) flag marked in the TCP header to acknowledge the receipt of the congestion indication. Note that the ECE and CWR flags in the TCP header are also used for the negotiation of ECN capability during connection establishment. See~\cite{ramak} for more details.

\begin{table}[h] 
\caption{Default ECN settings in major OSs}\label{tbl:osversion}
\begin{center}
\scalebox{0.9}{%
\begin{tabular}{|l|l|} \hline
OS and version                                  & Default ECN setting \\  \hline\hline
\multirow{2}{*}{Linux kernel 2.4.20 $\uparrow$}   & Enabled for incoming ECN\\ 
~                                               & connections  \\ \hline
\multirow{2}{*}{Mac OSX 10.11, iOS 9 $\uparrow$}         & Enabled for incoming ECN\\ 
~                                               & connections  \\  \hline
\multirow{2}{*}{Windows Server 2012 $\uparrow$}   & Enabled for both incomming \\ 
~                                               & and outgoing ECN connections  \\ \hline
\end{tabular}}
\end{center}
\end{table}

Table~\ref{tbl:osversion} shows default ECN settings in the current major OSs.
Most modern OSs use ECN and echo incoming ECN connections, but do not initiate the use of ECN on outgoing connections.
In other words, many networks and servers are ready for ECN. However, client
devices~\cite{enable-ecn} or applications~\cite{windows-ecn} need to enable ECN for it to be used.

\subsection{L4S}\label{subsec:lld-background}

L4S is a service architecture that uses ECN as an integral component to achieve high bandwidth and low latency for Internet applications and is currently undergoing standardization~\cite{ietf-tsvwg-l4s-arch-17}. The main idea behind L4S is to take advantage of the fact that the congestion signal in ECN isn't a degradation in the way that packet drops are, and thus congestion signals can be sent much more frequently in order to provide high-fidelity congestion information.  As mentioned above, L4S redefines the ECT(1) codepoint to indicate L4S-Capable Transport, and in the context of an L4S flow it redefines the CE codepoint to provide this fine-grained congestion feedback.  

L4S supports incremental deployment, via a classic congestion control response to packet drops, and via in-network isolation of classic traffic from L4S traffic.  L4S routers isolate these two types of traffic from one other so that the queuing latency caused by classic traffic doesn't impact L4S traffic, and so that the two types can each be provided their appropriate congestion signals. Two queuing mechanisms have been defined for this purpose.  One is referred to as Dual-Queue Coupled AQM, the other is an L4S-aware flow queuing approach. 

Dual-Queue Coupled AQM routers have two separate queues at the network bottleneck one for L4S traffic and one for classic traffic, where the  ECN field is used as an identifier for L4S packets to be distinguished from classic packets. The queue for L4S traffic has a shallower (or smaller) buffer size, which allows the L4S packets to experience very low queueing delay. Additionally, packets in the L4S queue are marked with CE in the ECN field (to notify the endpoints of impending congestion) as soon as they start building up in the queue (e.g. when the queue delay exceeds a low threshold of 500 µs or 1 ms). However, the queue for classic traffic has a larger buffer to maintain full utilization since the queue needs to be large enough to cope with large saw-tooth rate variations by a classic congestion control. In addition, despite the use of separate queues, the congestion signaling is coupled between the queues so that the two types of traffic share the bottleneck bandwidth in a fair manner. See~\cite{ietf-tsvwg-l4s-arch-17,ietf-tsvwg-aqm-dualq-coupled-23} for more details.

The L4S-aware flow queuing approach provides a separate queue for each individual flow based on a hash of the header 5-tuple, and then provides CE-marking of ECT(1) packets via a shallow queue delay threshold (e.g. 1 ms) while ECT(0) or Not-ECT packets are CE-marked or dropped (respectively) using a classic AQM algorithm.   

L4S is not limited to a certain type of network but can be implemented in different types of networks. However, the implementation of L4S in different networks poses different sets of obstacles, especially when L4S is implemented in 5G~\cite{brunello2021low, 5gecn2021erricsson} and WiFi~\cite{wifil4s}. 

\begin{figure*}[t]
     \centering
     \begin{subfigure}[Request and respond method~(\S\ref{sec:method1})]{
         \centering
         \includegraphics[height=4.35cm]{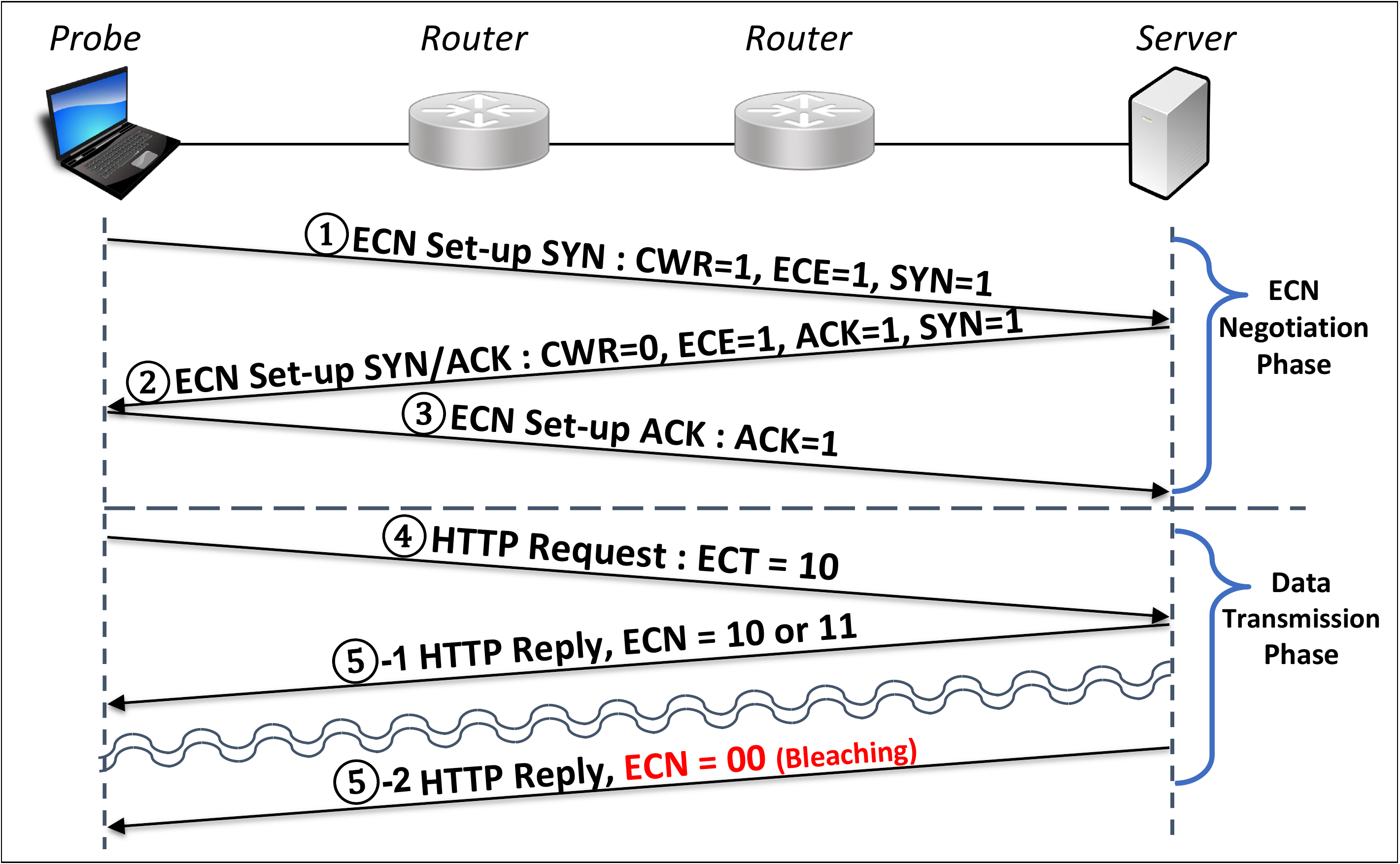}
         \label{fig:pathspider_method}
         }
     \end{subfigure}
     \begin{subfigure}[Traceroute method~(\S\ref{sec:method2})]{
         \centering
         \includegraphics[height=4.35cm]{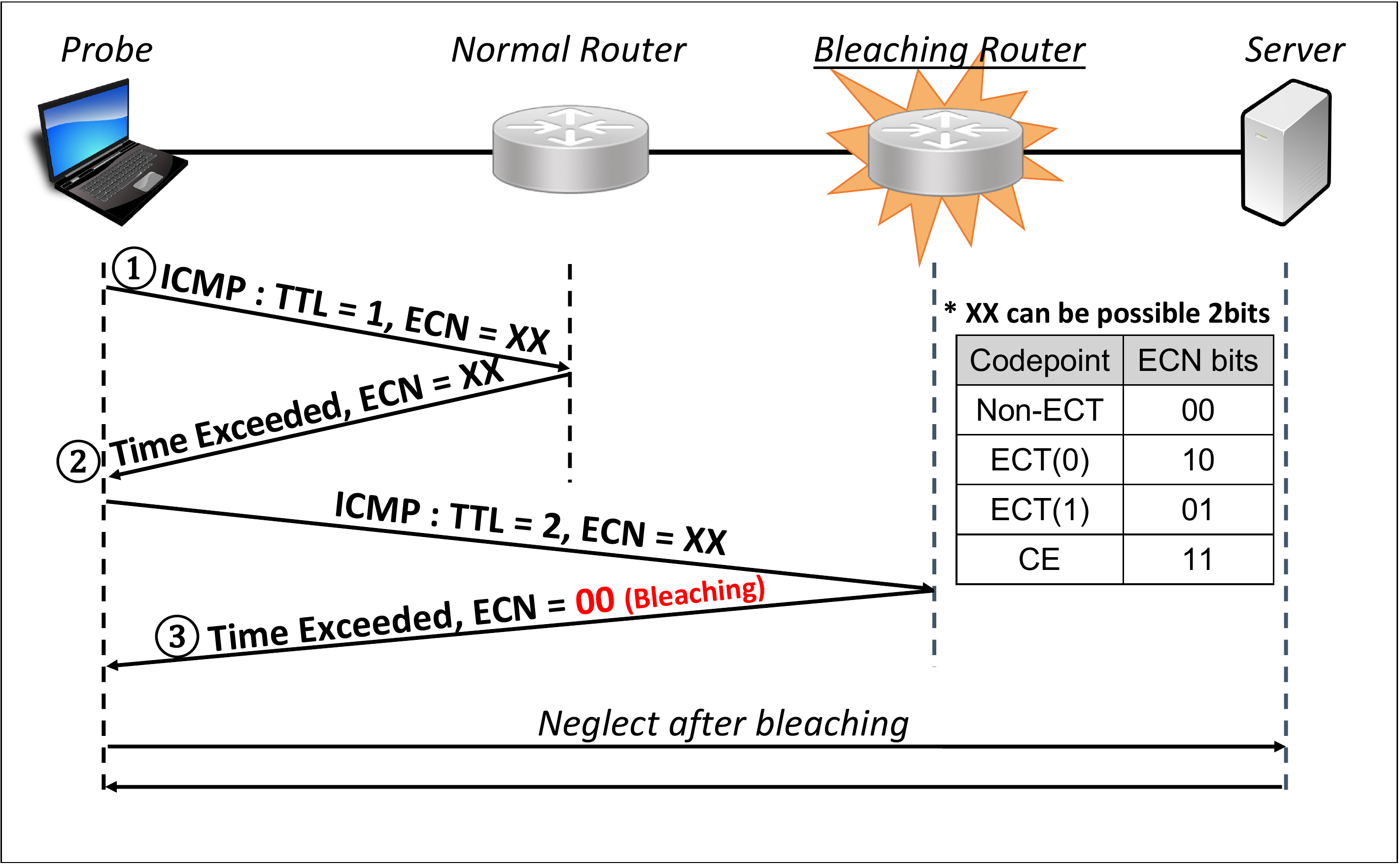}
        }
     \end{subfigure}
     \begin{subfigure}[Passive method~(\S\ref{sec:method3}) ]{
         \centering
         \includegraphics[height=4.35cm]{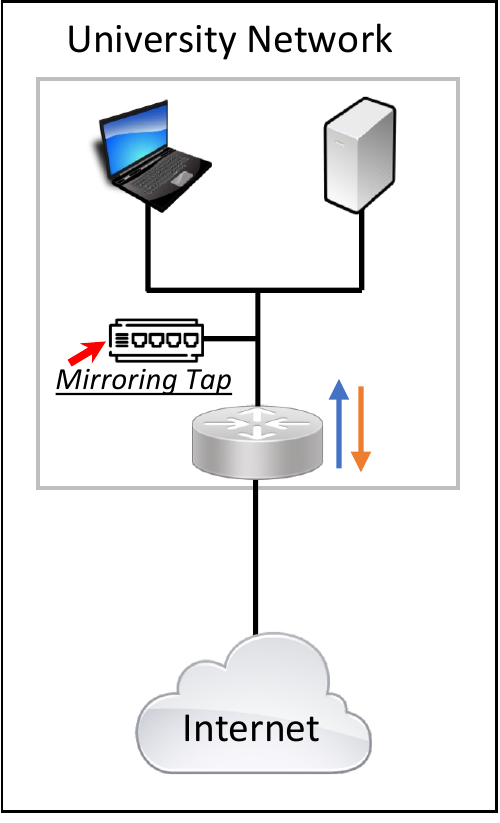}
        }
     \end{subfigure}
    \caption{\textmd{\textbf{(a) Request and respond method:} \textcircled{1} A probe starts ECN negotiation with a TCP SYN segment having the ECE and CRW flags set in the TCP header. \textcircled{2} The server responds with a SYN/ACK segment with ECE set. \textcircled{3} The probe sends an ACK segment to finalize the negotiation. \textcircled{4} The probe sends an HTTP request encapsulated in an IP packet with ECT(0) set. \textcircled{5}-1 In normal cases, the server sends back an ECN-marked IP packet. 
    \textcircled{5}-2 However, if ECN bleaching happens, the probe will receive packets that have the \emph{cleared} ECN field (\texttt{0b00}). 
    \textbf{(b) Traceroute method:} 
    \textcircled {1} A source probe sends ICMP echo request packets, each with incrementing TTL values, to a target server/probe. 
    \textcircled{2} When TTL expires, the router sends back an ICMP time exceeded error message to the source. 
    \textcircled{3} If a router bleaches the ECN field, the IP packet carrying the error message has \texttt{0b00} in the ECN field.
    \textbf{(c) Passive method:} We used a mirroring tap installed with the campus network to capture all traffic traversing the network.}}
\label{fig:setup}
\end{figure*}

\section{Measurement Methodology}\label{sec:method}

In this section, we explain the details of our methodology for active and passive measurements to examine the \emph{status quo} of the deployment and traversal of ECN in a wide range of wired and cellular networks. Specifically, we use a `request and respond' method and a \texttt{traceroute}-based method for active measurement to test or probe specific paths to measure ECN capability. To complement the active measurement results, we also passively collect and analyze a 24-hour trace of network traffic at a vantage point within a large University campus network to understand what fraction of traffic is ECN-enabled or how widely ECN is currently being used in practice.

\subsection{Active Measurement}

\subsubsection{Request and Respond Method}\label{sec:method1}

We use the active measurement tool called PATHspider~\cite{learmonth2016pathspider} for the request and response method to investigate the server-side deployment of ECN and the traversal of ECN over the paths to servers. There are two phases in this method, namely ECN negotiation phase and data transmission phase. 

First, the ECN negotiation phase allows us to confirm the ECN support of a target web server if the negotiation with the server succeeds. Specifically, a probe attempts to negotiate ECN capability with a targeted server while establishing a TCP session with the server. The probe first sends to the server a TCP SYN segment with ECE and CWR flags set in the TCP header. The server then sends back to the probe a SYN-ACK segment with ECE flag set, if the server supports ECN. Otherwise, the server just sends a plain SYN-ACK segment. After that, the server responds with an ACK to finalize the negotiation and establish a new session with the probe. It is worth noting that the ECN negotiation is invisible to routers in the path between the probe and the server since it is done at the transport layer. In other words, the routers do not affect the ECN negotiation even if there is a router in the path that bleaches the ECN field of IP packets.  

Second, the data transmission phase allows us to check whether ECN bleaching happens along the path from the probe to the server, assuming that the server supports ECN. The probe starts with sending an HTTP request to the server, where the IP packet carrying the HTTP request has ECT(0) marked in the ECN field. The server then sends an HTTP response back to the probe. The ECN field of the IP packet carrying (a part of) the HTTP response has either ECT(0) or CE marked, where ECT(0) is \texttt{0b10} and CE is \texttt{0b11}. While the former is for normal cases, the latter happens when a router in the path signals impending congestion by setting the ECN field of the packet with CE. However, if ECN bleaching happens at a router in the path (i.e., the router clears the ECN field to \texttt{0b00}), the IP packets received by the probe lose the ECN marking. Therefore, we can identify the presence of ECN bleaching by looking at the ECN field of incoming IP packets from the server. 

\subsubsection{Traceroute Method}\label{sec:method2}

To investigate whether the ECN-marked IP packets traverse the network without any illegitimate modifications, we use the \texttt{traceroute}-based method that has been used in the literature~\cite{bauer2011measuring,detal2013revealing,trammell2017tracking,mandalari2018measuring}. For a source probe and a target server/probe, this method relies on an ICMP time exceeded error message, which is generated at each router along the path to the target destination and sent back to the source. \texttt{Traceroute} sends a sequence of ICMP echo request packets whose IP packets have incremented time-to-live (TTL) values toward the IP address of the target destination. Each router discards an incoming IP packet when its TTL expires. In this case, the router also generates an ICMP time exceeded error message to the source. Upon receiving the error message at the probe, we can check whether there has been a change in the ECN field of the IP packet that carries the error message. Thus, in the \texttt{traceroute}-based method, we cycle through the ECN bits of the IP packets carrying ICMP echo request packets from \texttt{0b00} to \texttt{0b11} (recall the four ECN codepoints in Table~\ref{bg:l4s_ect}), while the TTL values of the packets increment. 

\subsubsection{Measurement Probes}
For the above active measurements, we use three types of probes, which are crowdsourced probes, cellular probes, and cloud probes.

\vspace{1mm}
\textbf{Crowdsourced probes.} We deployed the ECN measurement probes within the home and work ISP networks of 11 volunteers who participate in this study. The majority of the volunteers are located in the U.S., and the others are in Canada, Argentina, and Germany. Specifically, the 11 volunteers from four countries installed measurement probes in ten different home and work ISP networks, namely, Armstrong, Charter, Comcast, Cox, Rogers, Shaw, Telecom Argentina, Verizon business, and Vodafone. Many of the probes are deployed behind NAT/firewall in which case they only function as source probes for the \texttt{traceroute}-based method. Note that we did not use PATHspider with the crowdsourced probes since the volume of the traffic generated by PATHspider can be a burden to the volunteers.

\vspace{1mm}
\textbf{Cellular probes.}
We set up laptops with USB tethering as cellular probes to connect to cellular networks in order to examine ECN capability in cellular networks. We consider the cellular networks of six major US and South Korean carriers in this study, namely AT\&T, Verizon, T-mobile, KT, LGU+, and SKT. The cellular probes do not have public IP addresses to be connected from the outside of a network, so they only act as source probes for the request and respond method. 

\vspace{1mm}
\textbf{Cloud probes.}
We placed 37 cloud probes (or vantage points) in 33 geographic regions around the world and within the cloud servers operated by five different cloud service providers, namely, AWS, Azure, GCP, DigitalOcean, and Cloudlab. Cloudlab~\cite{cloudlab} is the only non-profit cloud service that provides high-performance computing and networks across several states in the U.S.
We set up a virtual machine (VM) server at each one of multiple cloud servers for each provider. 
The VM servers installed in multiple locations are to reflect the presence of multiple cloud servers across different geographical regions. In addition, each VM server has a public IP address, which allows it to be connected from other probes. 

It is worth noting that we would not get meaningful results if we only use cloud probes, as most of the primary cloud service providers, such as Amazon, Google, and Microsoft, have massive private WANs, which hide tenant traffic from the public Internet.

\subsection{Passive Measurement}\label{sec:method3}

While the above two active measurement methods allow us to examine the status quo of the ECN readiness in various types of networks, they cannot be used to reveal how widely ECN is currently being used in practice. To this end, we collected and analyzed a 24-hour trace of network traffic, whose data size is 10GB, at a vantage point within a large University campus network, from which we can see how widely ECN is currently being used among the traffic over the campus network.

We post-processed the traffic data as follows. We first obtained packet traces that only contain the IP and TCP/UDP headers of packets by stripping out their actual payloads. We next identified a packet trace per flow. The flow information that we collected includes the source and destination IP addresses and their country codes, port number, the ECN codepoint of the ECN field in the IP header, and the ECE and CWR flags in the TCP header, if the flow is a TCP flow. From the per-flow information, we were able to obtain the overall distribution of the ECN codepoints in the IP header as well as that of the ECE and CWR flags in the TCP header. We were also able to find the separate distributions of the ECN codepoints per port number. We finally investigated possible causes for the case when ECN(0) or ECN(1) is set in the ECN field of IP header while its upper transport protocol is not TCP.
\section{Measurement Results}
In this section, we closely examine the ECN field in the wild. 
We first present the ECN support results from our vantage points to public websites. We then use the \texttt{traceroute}-based method to understand how many Internet paths support ECN and pinpoint where ECN bleaching happens. 
Lastly, we examine the overall usage of ECNs from traffic traces collected on a University campus.

\subsection{ECN-enabled Public Websites}\label{results_1}
\subsubsection{Methodology.}\label{results_1:method} 
To consider CDN deployment of popular websites listed as Alexa 100K website domains~\cite{alexa}, we chose 16 different vantage points out of 25. We then obtained 404,382 unique IP addresses for these websites by removing duplicate ones. We use the request and response method using PATHspider to check if ECN is supported from each vantage point to each website, where our probe in each vantage point sends an HTTP request and checks the response. We ensure that a middlebox or the network where each vantage point is hosted is not mangling TCP and IP headers.

\begin{figure}[h]
    \centering
    \includegraphics[width=0.95\linewidth]{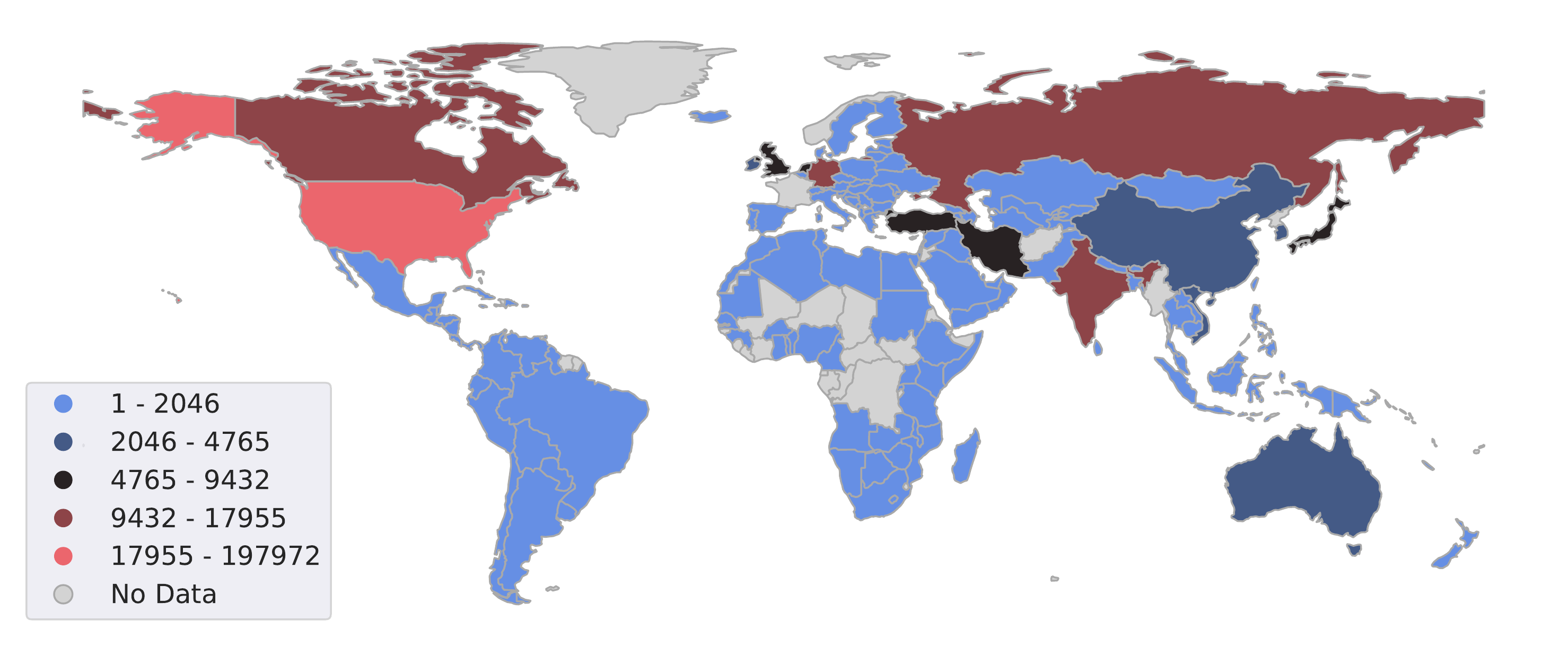}
    \caption{\textbf{Geographic locations of tested Alexa 100K web servers.} \textmd{Since popular domains provide their service from geographically distributed clouds, the total number of unique IPs \textbf{404,382} is much more than the number of listed domains. \textbf{U.S.} accounts for most of the IPs by \textbf{53.6\%}, followed by Germany (4.8\%), Russia (4.3\%), and Canada (3.8\%).}}
\label{fig:location_continents}
\end{figure}

\textbf{Web server locations}. We are interested in understanding where these websites are geographically located. We use \texttt{whois} command to locate each IP address. Figure~\ref{fig:location_continents} shows the locations of Alexa 100K websites by using the location information (i.e., City, StateProv, and Country) from each \texttt{whois} query. The majority of Alexa 100K websites are located in the U.S. (56\%), followed by Germany (4.8\%), Russia (4.3\%), and Canada (3.8\%).

\begin{figure}[t]
    \centering
    \includegraphics[width=0.8\linewidth]{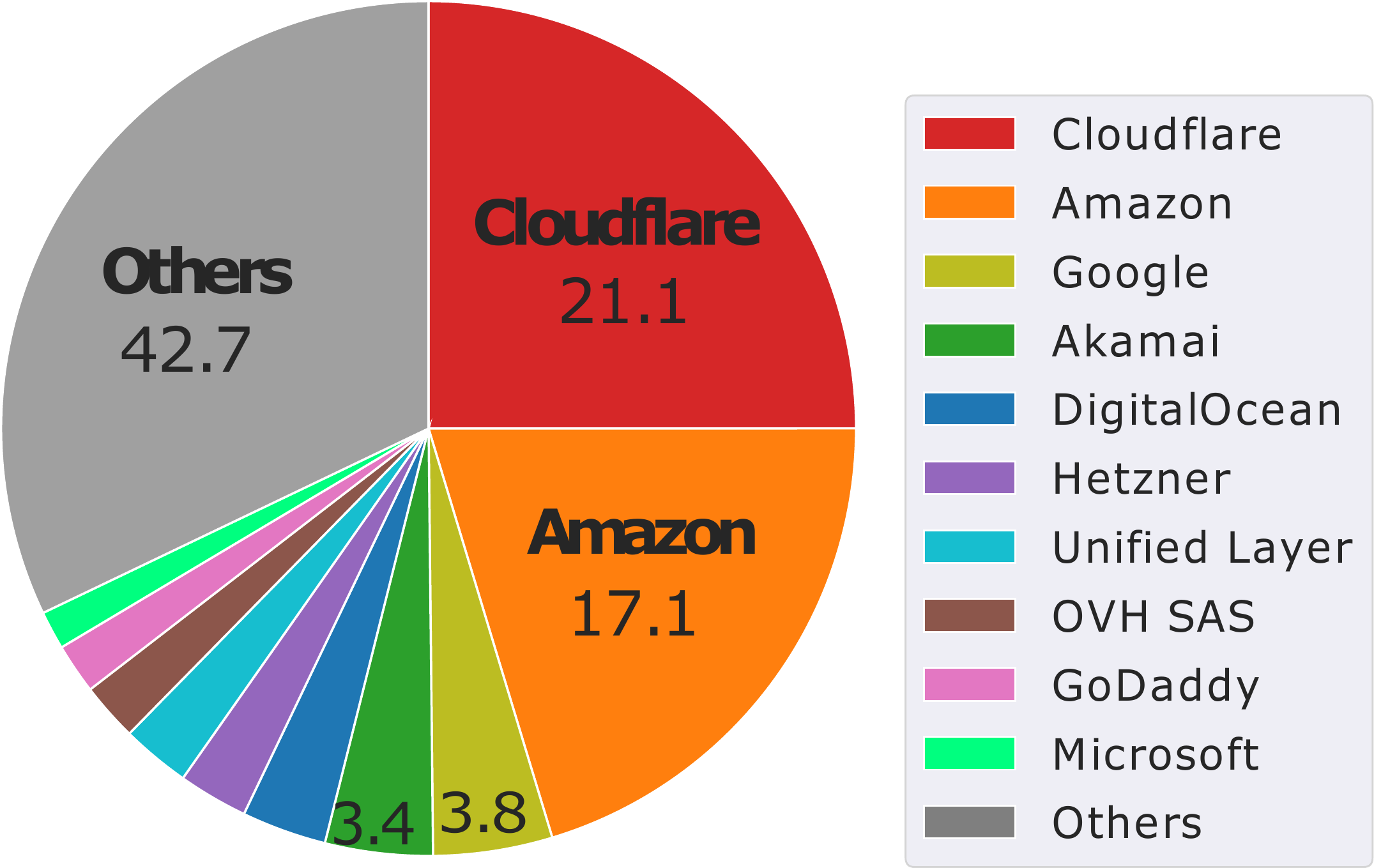}
    \caption{Top 10 web-server hosting services of Alexa 100K web servers.
\textmd{The total of 10,950 unique Web-server hosting providers are identified for Alexa 100K websites. Among 404,382 IPs, Cloudflare (74,510) and Amazon (63,285) show the highest proportion.}}
\label{fig:location_isps}
\end{figure}

\vspace{1mm}
\textbf{Web server hosting providers.} 
We are also interested in service providers hosting these websites. We classify each IP address based on the service providers' names using the \texttt{whois} response. Figure~\ref{fig:location_isps} shows  the top 10 providers for Alexa 100K websites. We obtained around 10,950 providers serving these websites. We found that many web servers are hosted in major data centers in the U.S. with Cloudflare and Amazon being the top 2 providers, serving 21.1\% and 17.1\%, respectively. 

\begin{figure}[t]
    \centering
    \includegraphics[width=0.85\linewidth]{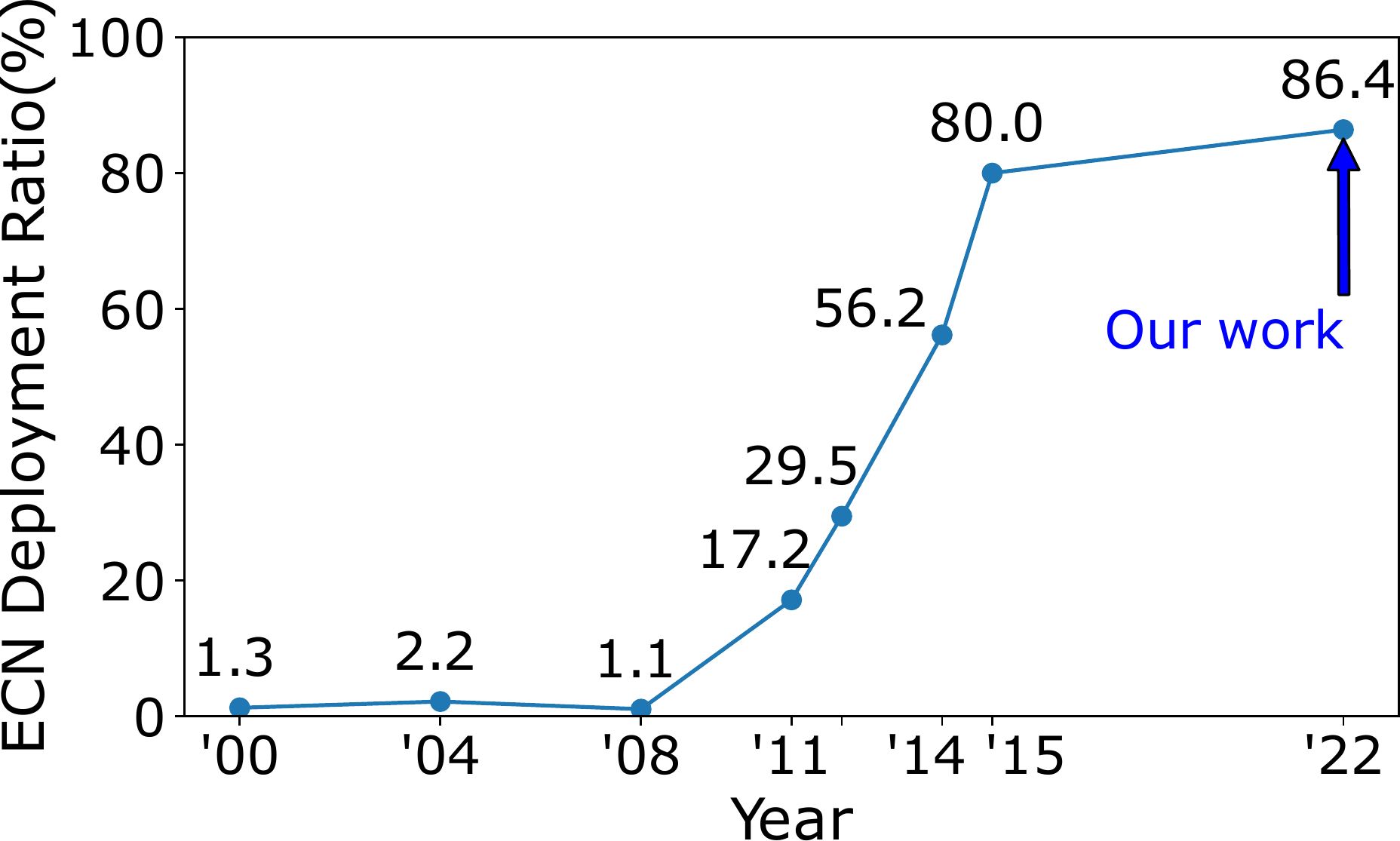}
    \caption{ECN deployment status in chronological order.
    \textmd{ECN deployment has rapidly increased to 80\% in 2015's measurement study. In our measurement study, ECN deployment has risen to 86.4\%, following the trend shown in 2015's measurement study showing 80\%.}
    }
\label{fig:setup_codepoint}
\end{figure}

\subsubsection{ECN deployment status on the servers.}\label{results_1:overview} 
Figure~\ref{fig:setup_codepoint} shows the ECN deployment ratio over the last two decades in chronological order. Measurement studies~\cite{padhye2001identifying,medina2005measuring,langley2008probing} before 2010 reported an almost negligible number of ECN-enabled web servers. The study in 2015~\cite{mcquistin2015explicit} reported that ECN deployment had rapidly increased to 80\%. Our measurement study shows that the percentage of ECN-supported web servers is now 86.4\%, following the trend shown in the 2015 study. These servers offer successful ECN negotiation to the client during TCP session establishment; this percentage still includes the cases where a TCP session with successful ECN negotiation is bleached in IP packets by some intermediate routers.

Table~\ref{result:total_ecn_endis} shows the details of this measurement results. Among 349,188 (85.4\%) ECN-enabled Web servers completing successful ECN negotiation in TCP session establishment, 37,798 (10.8\%) hosts show ECN bleaching in IP packets; we conjecture that these paths have ECN bleaching points between the vantage points and these Web servers. The 0.2\% percentage of CE marking on ECN-enabled connections matches the results by Apple in 2017~\cite{Padma2027apple} that the percentage of CE marking in the packets between two ECN-enabled devices in the U.S. shows 0.2\% while it varies from country to country. 

\begin{table}[t] 
\begin{center}
\caption{ECN-enabled web servers obtained from one vantage point to the list of public websites (404,382 IPs).}\label{result:total_ecn_endis}
\begin{tabular}{| l |r r | r r |} \hline
Codepoint             & Enabled   &(pct)    & Disabled  &(pct) \\ \hline\hline
Total                 & 349,188   &(86.4\%) & 55,194    &(13.6\%) \\ \hline\hline
\texttt{00}: Not-ECT  & 37,798    &(10.8\%) & 55,141    &(99.9\%) \\ \hline
\texttt{10}: ECT(0)   & 259,634   &(74.4\%) & 48        &(0.09\%) \\  \hline
\texttt{01}: ECT(1)   & 50,898    &(14.6\%) & 3         &(0.01\%) \\ \hline
\texttt{11}: CE       & 858       &(0.2\%)  & 2         &(0.004\%) \\ \hline
\end{tabular}
\end{center}
\end{table}

\subsubsection{ECN-enabled paths within or across continents}
\begin{figure}[h]
     \centering
     \begin{subfigure}[ECN-enabled path ratio]{
         \centering
         \includegraphics[width=0.469\linewidth]{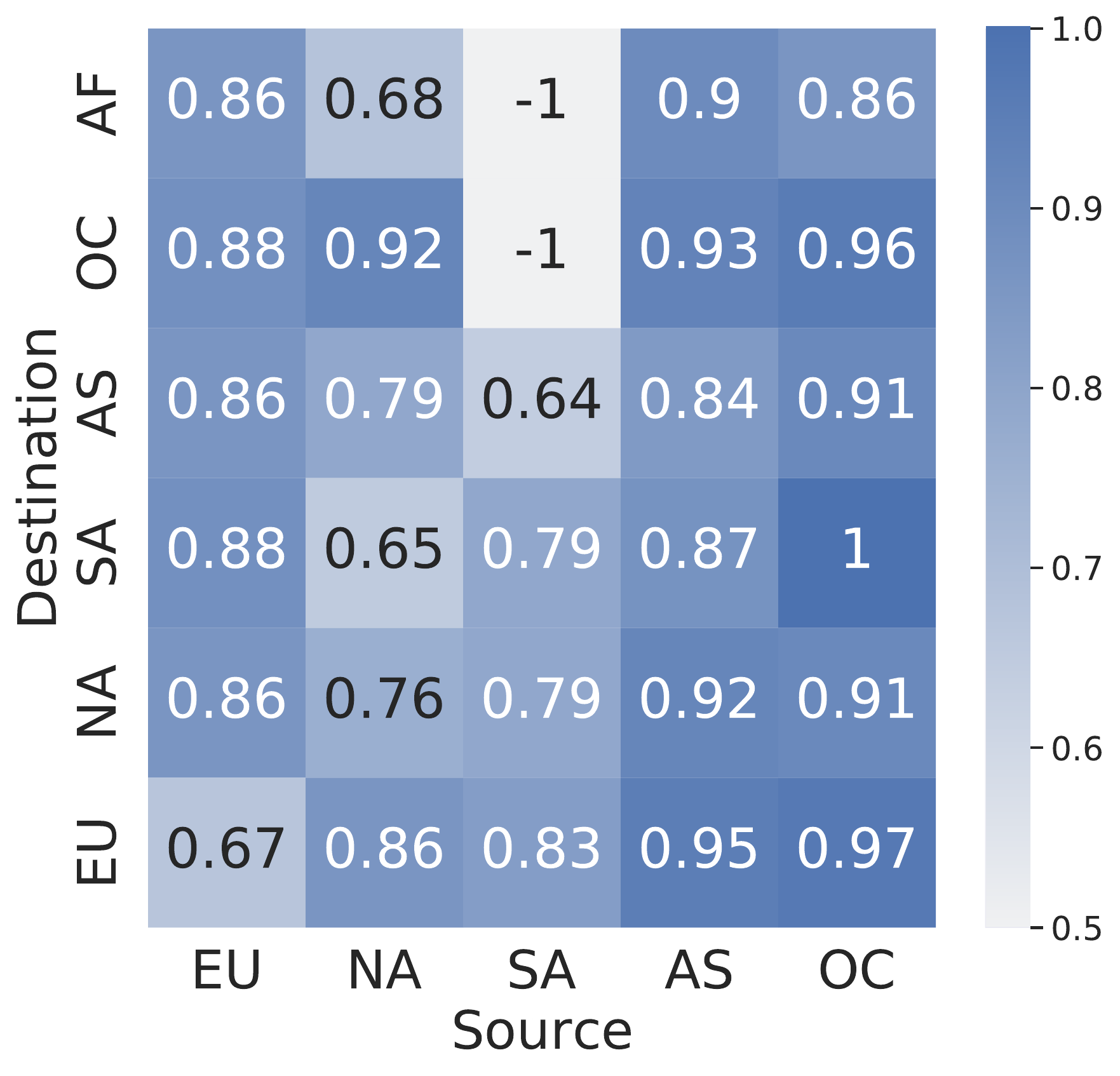}
         \label{fig:figure_ECN_enabled}
         }
     \end{subfigure}
     \begin{subfigure}[ECN-bleaching path ratio]{
        \centering
        \includegraphics[width=0.469\linewidth]{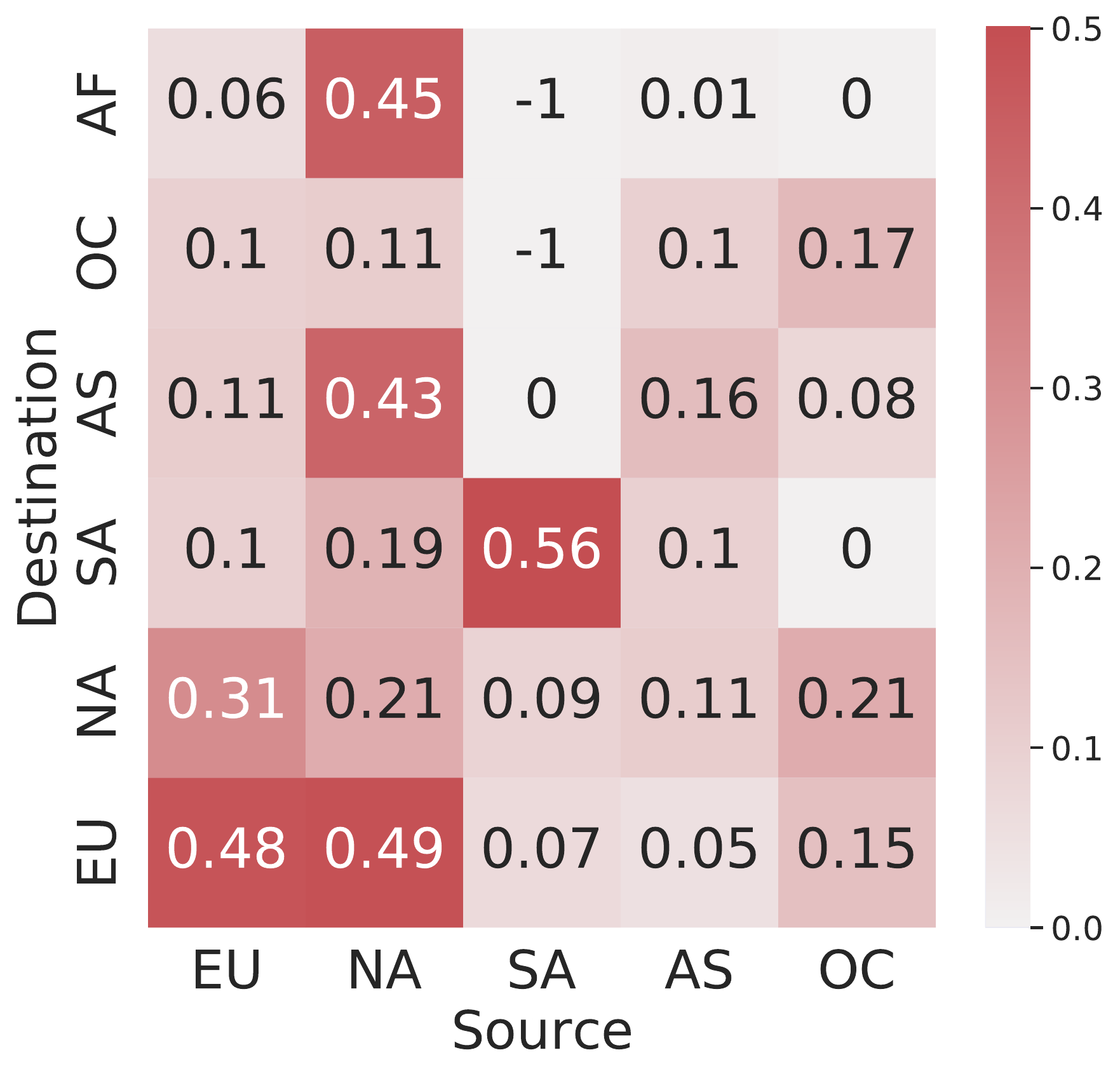}
        \label{fig:figure_ECN_bleaching}
        }
     \end{subfigure}
     \caption{ECN-enabled and ECN-bleaching path ratios within and across continents. 
          \textmd{$-1$ represents no results are available. ECN-enabled paths are less correlated with the geographical locations of clients and web servers (a). Europe (EU) and South America (SA) show the highest bleaching ratio within the same continent. The requests sent from vantage points in North America (NA) show higher ECN bleaching rates (over 40\%).}
     }
\label{fig:Figure4}
\end{figure}

By sending a request from the vantage points hosted in cloud providers where no local ECN bleaching is observed to public websites, we measure the percentage of ECN-enabled paths showing successful ECN negotiation in TCP sessions. Figure~\ref{fig:figure_ECN_enabled} visualizes ECN-enabled paths within or across continents. $-1$ indicates no available measurement data. Interestingly, no particular region supports ECN better than the others. We conjecture that ECN-enabled paths depend largely on the OS version and settings of hosting web servers which may not correlate with geographical locations. If we average for each destination continent, the ECN-enabled path ratio becomes close to 86.4\% of the ECN deployment ratio in Figure~\ref{fig:setup_codepoint}.

\subsubsection{ECN-bleached paths within or across continents}
From each ECN-enabled path of successful ECN negotiation in TCP sessions (Figure~\ref{fig:figure_ECN_enabled}), we show the ECN bleaching ratio where ECN bits are wiped in IP packets (Figure~\ref{fig:figure_ECN_bleaching}).  
Europe (EU) and South America (SA) show the highest bleaching ratio within the same continent. North America (NA) was the highest ECN bleaching source among the other continents. The Africa (AF) websites show relatively low bleaching rates except for NA-originated traffic, while Oceania (OC) websites show the lowest ECN bleaching rates. These results illustrate that the ECN bleaching points are located primarily in North and South America and Europe. 

\subsubsection{ECN bleaching from different access networks}
We check the impact of different access networks when clients located in each access network connects to public websites. 
Table~\ref{tbl:pathspider_results_per_cellularprovider} shows the summary of ECN negotiation and bleaching percentages for different types of access networks and providers. In wired access networks, Comcast shows a 51.35\% ECN bleaching percentage among the ECN-enabled paths with successful ECN negotiation (86.8\%). In contrast, the other providers show around 10\% ECN bleaching percentage, which matches the overall ECN deployment percentage in Table~\ref{result:total_ecn_endis}.

In cellular networks, AT\&T shows ECN negotiation responses come from the middleboxes at the TCP connection establishment phase, not by the destination web servers. We confirmed this behavior by sending an ECN negotiation packet to ECN-disabled web servers. Surprisingly, the client successfully negotiates ECN with any web server (99.8\%)\footnote{We confirm that this is not because of the USB tethering.}, indicating the presence of performance-enhancing proxies (PEPs) in the network. On the other hand, SKT in South Korea bleaches 100\% of ECN-negotiated sessions; it wipes ECN bits in all IP headers.

\begin{table}[t]
\small
\caption{ECN bleaching from different access networks in the U.S. and South Korea.} \label{tbl:pathspider_results_per_cellularprovider}
\begin{tabular}{|c||c|c|c|c|}  
\hline  
Type                        & Provider  & location  & Negotiation   & Bleaching     \\ 
                            &           &           & Percentage    & Percentage    \\\hline\hline  
\multirow{4}{*}{Wired}      & Comcast   &  US      & 86.8\%  &    \textbf{51.35}\%      \\  
                            & LG U+        &  S. Korea          & 89.1\%  &    9.82\%  \\  
                            & University A      &  US      & 86.4\%  &    13.6\% \\  
                            & University B     &  S. Korea      & 92.3\%  &    8.93\% \\  \hline  

\multirow{6}{*}{Cellular}   &   AT\&T    &  US      &   \textbf{99.8\%}  &   28.3\%   \\  
~                           &   Tmobile   &  US      &   86.7\%      &    5.1\%   \\  
~                           &   Verizon   &  US      &  61.5\%      &    3.1\%   \\ 
~                           &   SKT   &  S. Korea      &     91.7\%  &   \textbf{100\%}  \\ 
~                           &   LG U+   &  S. Korea      &  89.5\% &   9.05\%   \\ 
~                           &   KT   &  S. Korea      &    92.3\%   &    7.64\% \\\hline  
\end{tabular}  
\end{table}  

\subsubsection{ECN bleaching from different cloud providers}
We now check if cloud providers support ECN when clients in their cloud connect to public websites. 
Table~\ref{tbl:table6} shows the results. Interestingly, clients in Azure failed 100\% in ECN negotiation. We conjecture that PEPs in its network change the ECN field without following the standards. We confirmed the alteration at the middleboxes with additional testing with our own servers with ECN enabled and disabled, respectively. Another key observation is that GCP bleaches $100\%$ in IP headers even after successful ECN negotiation in TCP headers. We conjecture that Google removes ECN in IP packets to use it internally in their datacenters.

\begin{table}[t]
\small
\caption{ECN bleaching from different cloud providers.}\label{tbl:table6}
\begin{tabular}{|c||c|c|c|}  
\hline  
Cloud Service        & Location  & Negotiation   & Bleaching \\ 
Provider          &           & Percentage    & Percentage  \\\hline   \hline  
\multirow{4}{*}{AWS}&Frankfurt&60.2\% &39.41\% \\
~&Stockholm&85.24\% &9.31\% \\
~&SaoPaulo&75.81\% &19.97\% \\
~&Paris&60.66\% &41.11\% \\
\hline
\multirow{4}{*}{Azure}&Norway&\textbf{0\%} &-\\
~&NorthEU&\textbf{0\%} & - \\
~&UK&\textbf{0\%} &- \\
~&East Asia&\textbf{0\%} & -\\
\hline
\multirow{4}{*}{Cloudlab}&Wisconsin&97.8\% & 15.3\%\\
~&Utah& 93.7\%& 9.17\% \\
~&Clemson& 95.9\%& 5.61\%\\
~&Massachusetts& 91.4\%& 7.87\%\\

\hline
\multirow{8}{*}{DigitalOcean}&New York&88.46\% &40.86\% \\
~&San Francisco&92.59\% &29.81\% \\
~&Singapore&93.28\% &16.85\% \\
~&London&88.23\% &39.2\% \\
~&Frankfurt&87.55\% &35.48\% \\
~&Toronto&81.8\% &38.54\% \\
~&Bangalore&95.09\% &4.71\% \\\hline 
\multirow{2}{*}{GCP}&  Virginia   &   94.1\% &   \textbf{100\%}    \\
~&California&93.9\% &\textbf{100\%} \\
\hline 

\end{tabular}  
\end{table}  

\subsection{ECN-supported Internet Paths}\label{result:method_2}
\subsubsection{Methodology}
The trends in Subsection~\ref{results_1} highlight that the ECN deployment ratio in public websites is reaching 90\% and is already well used on the Internet, while the network bleaches 10\% of those ECN-enabled websites. To understand and pinpoint where in the network ECN fields are being altered or bleached, we use the \texttt{traceroute}-based method mentioned in Section~\ref{sec:method2} towards each ECN-enabled public website. We further categorize the types of ECN violations and seek potential causes of ECN bleaching.

\subsubsection{ECN deployment status in the network}
Table~\ref{result:ecntr_violation} shows the number of ECN bleaching network paths and their IP addresses. Among 534,077 paths we tested, which exclude five providers that bleach the ECN field as a matter of policy, 22,305 (4.17\%) paths bleach the ECN field. This corresponds to 1,112 bleaching router IPs among 129,252 IPs. We further divided ECN violation cases into three cases: (1) bleaching all ECN bits into zeros, (2) ECT(0) becomes ECT(1), and (3) CE becomes either ECT(1) or ECT(0). As we expected, more than 99\% of ECN violation cases fall into Case 1 of zeroing out all ECN bits.

\begin{table}[t] 
\renewcommand{\arraystretch}{1.2}
\begin{center}
\caption{Total number of ECN violating paths and bleaching IP addresses obtained from the traceroute method.}\label{result:ecntr_violation}

\scalebox{0.95}{%
\begin{tabular}{| l  l |  r r|r r|}
\hline
Input  & \textrightarrow \hspace{0.1cm} Output & \# of Path &(pct) & \# of IP &(pct)  \\
\hline\hline
\# of & Total Case& 534,077 & & 129,252 & \\ 
\hline\hline
\# of & Violation & 22,305 &(4.17\%) & 1,112 &(0.8\%) \\ 
\hline
\hspace{0.1cm}\texttt{Any} &\textrightarrow  \hspace{0.1cm} \texttt{00} & 22,111 &(99.1\%)& 1,070 &(96\%) \\
\hspace{0.1cm}\texttt{10} &\textrightarrow \hspace{0.1cm} \texttt{01} & 171 &(0.76\%)& 33 &(2.9\%)\\ 
\hspace{0.1cm}\texttt{11} &\textrightarrow \hspace{0.1cm} \texttt{10} or \texttt{01} & 24 &(0.1\%) & 9 &(\%)\\ 
\hline
\end{tabular}}
\end{center}
\end{table}

\subsubsection{ECN bleaching IPs and locations.}
To understand where those bleaching IP addresses are located, we use the whois command to locate each IP address. Table~\ref{result:totalnumber} shows the geographical distribution of those bleaching points. Europe and South/North America have proportionally higher bleaching points than the others. 

\begin{figure}[t]
    \centering
    \includegraphics[width=0.95\linewidth]{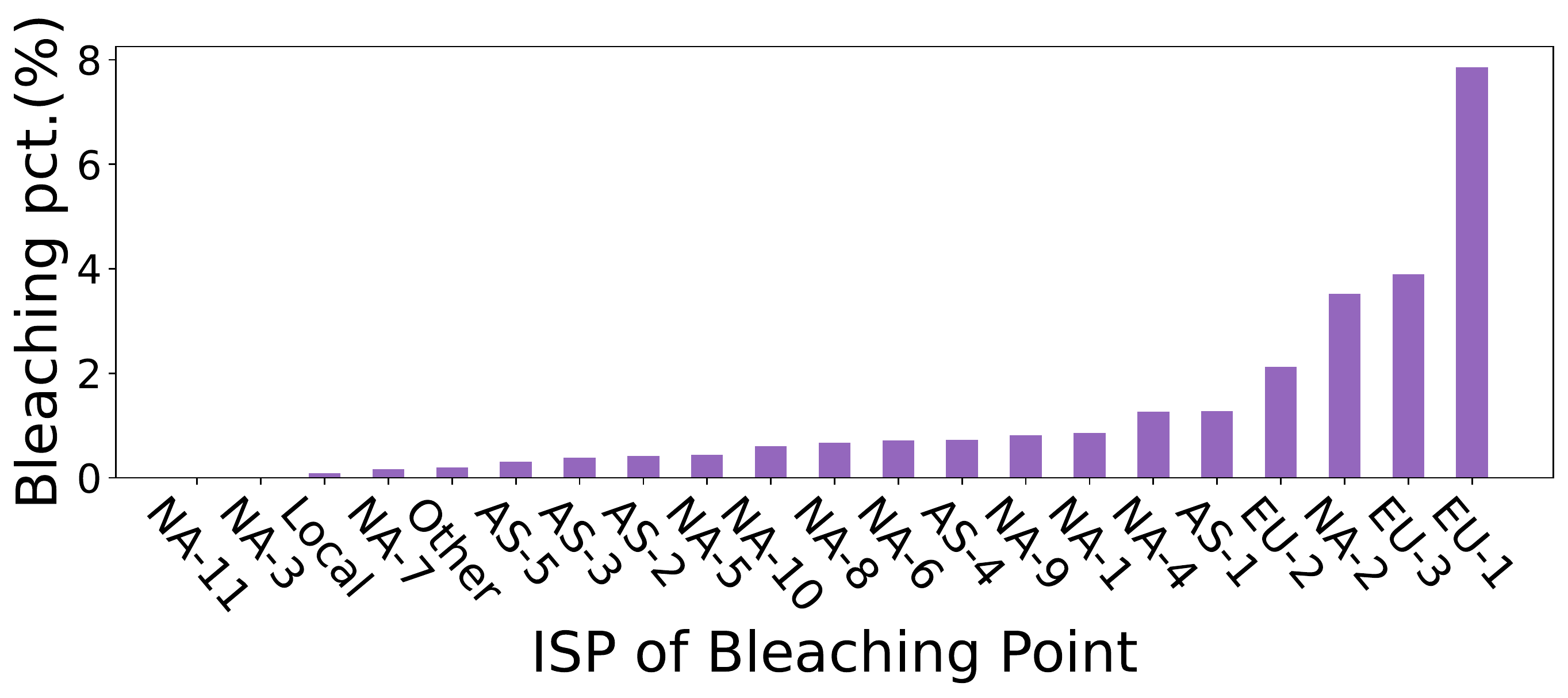}
    \caption{\textbf{ISPs of bleaching points.} 
    \textmd{The X-axis represents the ISP name of the bleaching IP addresses. From 1,112 bleaching IP addresses, ISPs in the EU have the highest ECN bleaching percentage.}}
\label{fig:Figure_bpctperisp}
\end{figure}

As we are interested in understanding which ISPs have those bleaching IP addresses, we compute the bleaching percentage for each ISP: how many ECN bleaching routers belong to each ISP.   
Figure~\ref{fig:Figure_bpctperisp} shows the percentage of bleaching routers belonging to each ISP. Due to its sensitivity, we anonymized the names of ISPs. Overall, three ISPs in Europe ranked 1, 2, and 4 in the bleaching percentage, matching the ECN-bleaching ratio obtained for public websites in Figure~\ref{fig:Figure4}.

\begin{table}[t] 
\begin{center}
\caption{Geographic locations of bleaching IPs (BIPs).} \label{result:totalnumber}
\scalebox{0.95}{%
\begin{tabular}{| c || r  | r || r |}
\hline
Continents & Total IP & \# of BIP & Percentage  \\
\hline\hline
Total&129,252 &1,112     &   0.08\%    \\\hline
\hline
North America & 44,617     &431 &0.97\%  \\\hline
Europe& 20,525    &513 &2.5\%    \\\hline
Asia & 29,171     &124 &0.43\%    \\\hline
South America& 1,900  &18 &0.95\%     \\\hline
Oceania& 1,057     &4 &0.38\%      \\\hline
Africa & 506      &0 &0.00\%      \\\hline
Unknown& 1,476   &22 &1.56\%     \\\hline
\end{tabular}}
\end{center}
\end{table}

\subsubsection{ECN bleaching network paths per source ISP}
By leveraging our probes in each ISP, we obtain the percentage of ECN bleaching paths among the total number of paths initiated from each ISP. Table~\ref{tbl:bleaching_pct_per_isp} shows the results. While it is hard to generalize as we only had a few probes in each ISP, traffic initiated from probes installed in Rogers and Vodafone wired networks and Google Cloud show 100\% ECN bleaching. 

\begin{table}[t]
\small
\caption{The percentage of bleaching network paths per source ISP. \textmd{}} \label{tbl:bleaching_pct_per_isp}
\begin{tabular}{|c||c|c|c|}  
\hline  
Type                        & Provider  & location  & Bleaching     \\ 
                            &           &           &Percentage    \\\hline\hline  
\multirow{9}{*}{Wired}      & Armstrong     &  US      & 69.73\%  \\  
                            & Charter  &  US          & 53.2\%   \\  
                            & Comcast   &  US      & 3.58\%     \\  
                            & Cox  &  US          & 38.2\%   \\  
                            & Rogers  &  US          & \textbf{100\%}   \\  
                            & Shaw     &  US     & 19.14\%  \\
                            & Telecom Argentina	     &  Argentina      & 16.57\%  \\
                            & University A      &  US      & 9.41\%  \\  
                            & Vodafone     &  Germany      & \textbf{100\%}  \\ \hline  
\multirow{3}{*}{Cellular}   &   AT\&T    &  US      &  48.2\%    \\  
~                           &   Tmobile   &  US      &   5.63\%         \\  
~                           &   Verizon   &  US      &  11.06\%         \\ \hline  

\multirow{4}{*}{Cloud}      &   AWS &  16 Locations   &    2.69\%       \\  
~                           &   Google& 10 Locations   &   \textbf{100\%}    \\  
~                           &   DigitalOcean& 8 Locations &  5.38\%   \\
~                           &   Cloudlab&Wisconsin  &  16.53\%   \\
~                           &   Cloudlab&Clemson  &  5.38\%   \\
~                           &   Cloudlab&Utah     &   2.69\%    \\  \hline 
\end{tabular}  
\end{table}  

\subsubsection{ECN bleaching hops.}
\begin{figure}[t]
    \centering
    \includegraphics[width=0.7\linewidth]{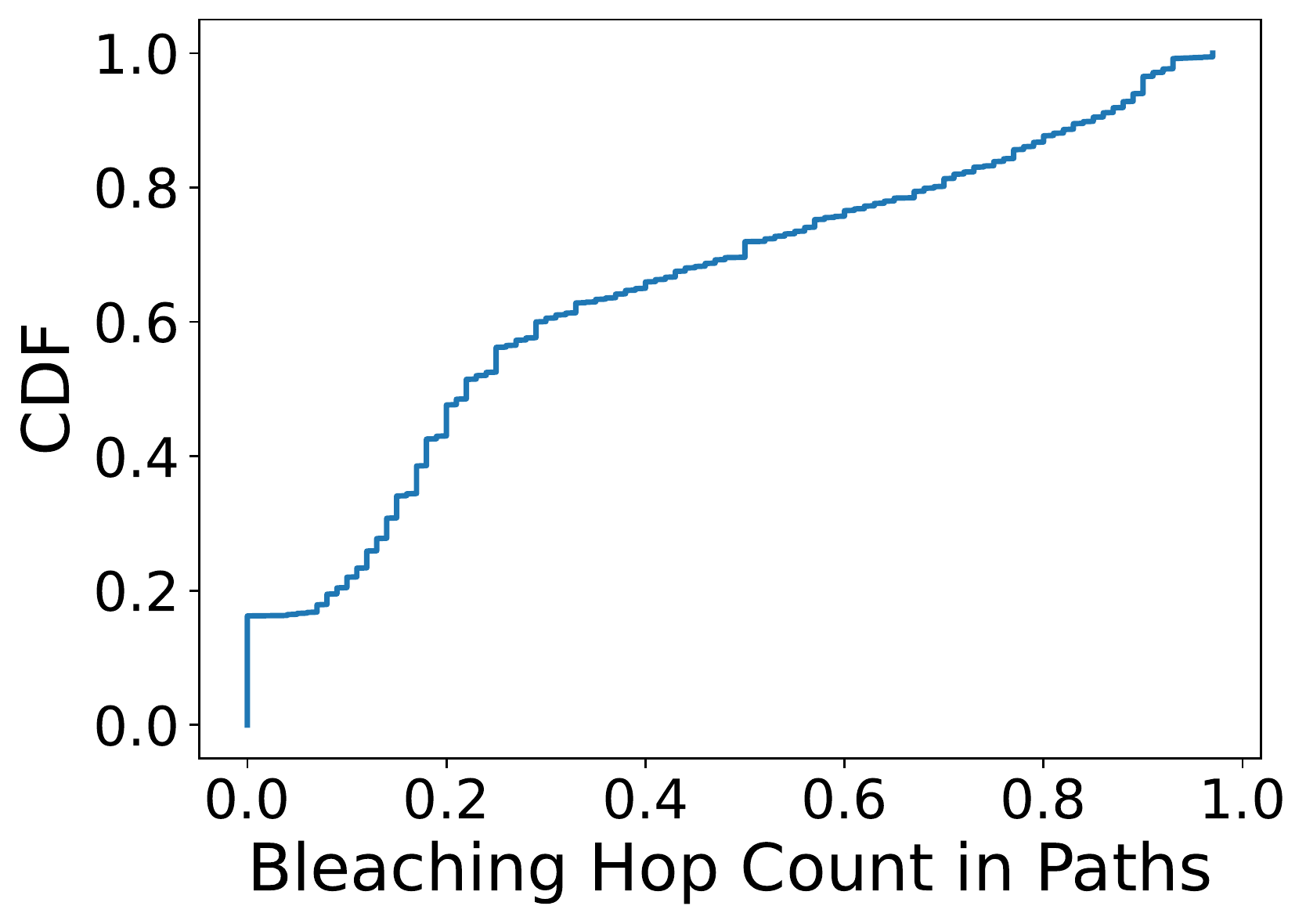}
    \caption{CDF analysis of bleaching distance in the normalized network path. The percentages of ECN bleaching were observed at relative hops in the paths. }
\label{fig:result_hops}
\end{figure}

The bleaching hop number represents where ECN bleaching occurs in the network path.
Figure~\ref{fig:result_hops} shows the percentages of ECN bleaching observed at relative hops in the paths. We divided a hop number, where we find ECN bleaching, from the total number of hops until the destination. 
Based on our results, we found that a large portion of bleaching happens in the access network. Also, if ECN bleaching occurs in the access network, the bleaching hop number is no longer than seven hops.
The overall bleaching hops are well distributed, but $50\%$ of ECN bleaching occurs in $30\%$ of the total distance between the source and destination. 

\subsubsection{ECN bleaching router signatures.}
\begin{figure}[t]
    \centering
    \includegraphics[width=0.7\linewidth]{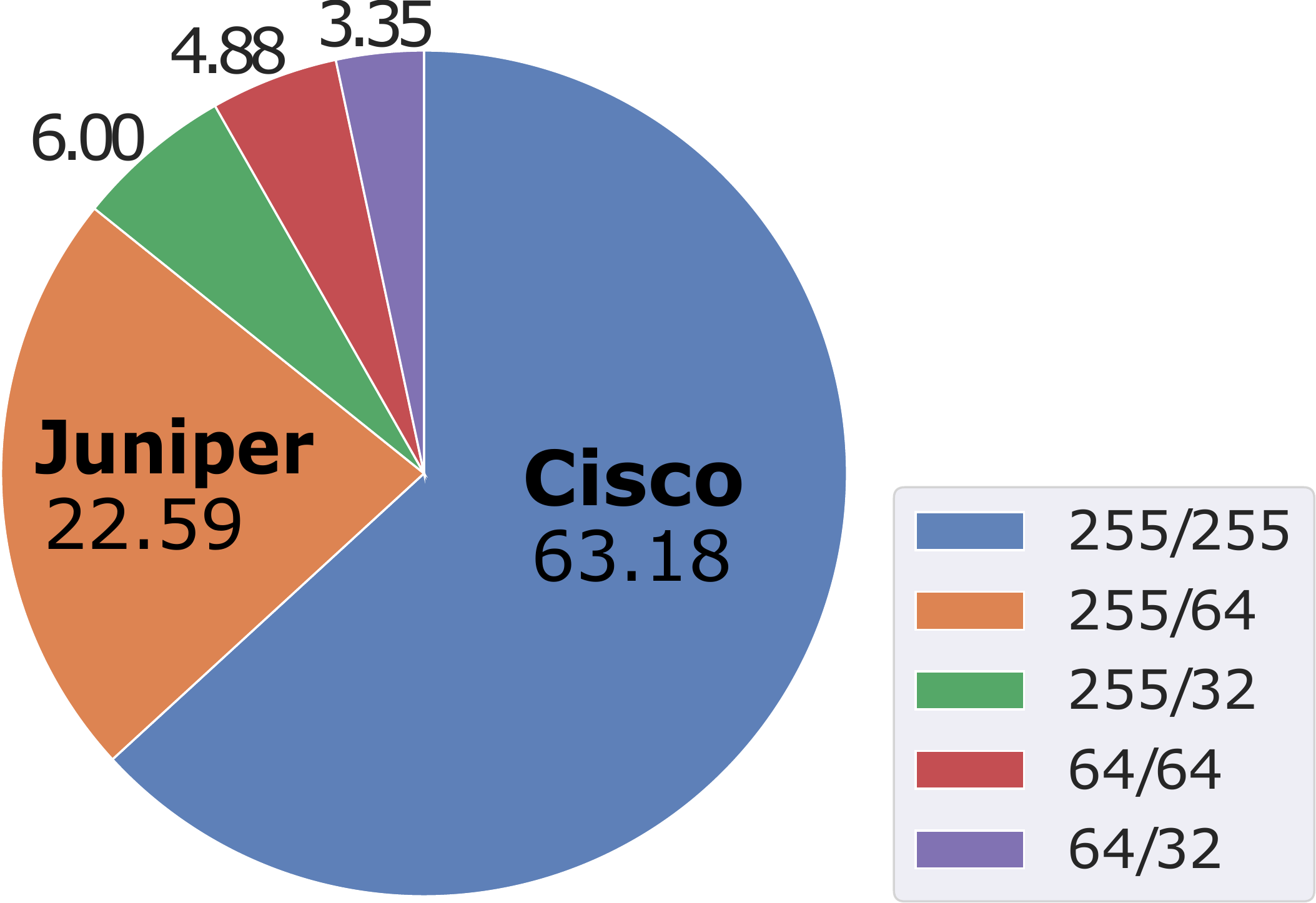}
    \caption{Vendor distribution of routers bleaching ECN. 
    \textmd{We adapt the technique described in \cite{vanaubel2013network}, with which we can identify a router as Cisco, Juniper or Others by comparing the IP TTL of the response to an ICMP echo-request and to a Time exceeded message.}}
\label{fig:ttl}
\end{figure}

\vspace{1mm}
We fingerprinted the routers with the list of bleaching IPs.
The fingerprint technique we used is from \cite{vanaubel2013network}, identifying the vendor of a router based on the response TTL from ICMP echo-request and Time exceeded message (e.g., 255/64).
The IP packet header contains a time-to-live (TTL) field used to avoid packets looping forever when a routing loop occurs. 
This 8-bit field is set by the originating host/router to an initial value that is usually and nearly always a power of 2 in the list 32 (or 30), 64, 128, and 255. The TTL field is decremented by one at each intermediate node along the packet's path. 
Figure~\ref{fig:ttl} shows the distribution of routers bleaching ECN. As we expected, many bleaching routers are from Cisco and Juniper. 

\subsection{Usage of ECN from Traffic Traces}

\begin{table*}[ht]
\caption{Overall results from passive measurement. \textmd{ECN is negotiated during TCP connection establishment. Note that the ECN percentage in TCP counts only within successful ECN negotiation flows. For UDP, we only collect the number of UDP flows with ECN codepoints since UDP does not have ECN negotiation standards.} }\label{result:passive_results}
\begin{tabular}{|c||c|c|r||r|r|r||r|r|r|r|}  
\hline  
                    &   &   &  &\multicolumn{3}{c||}{ECN Negotiation} &\multicolumn{4}{c|}{Within ECN Negotiation Succeeded} \\ 
Type                    &  Port &  Protocol & Flow Pct & \multicolumn{1}{c}{Succeeded}   & \multicolumn{1}{c}{Failed}    &\multicolumn{1}{c||}{No Attempt}  &\multicolumn{1}{c}{Non-ECT }  &\multicolumn{1}{c}{ECT(1)} &	 \multicolumn{1}{c}{ECT(0)}  & \multicolumn{1}{c|}{CE} \\ 
                        &           &        & & \multicolumn{3}{c||}{} &  \multicolumn{1}{c}{(\texttt{0b00})}  & \multicolumn{1}{c}{(\texttt{0b02})} &\multicolumn{1}{c}{(\texttt{0b10})}&\multicolumn{1}{c|}{(\texttt{0b11})} \\
                        \hline  \hline 
\multirow{11}{*}{TCP}& \multicolumn{2}{c|}{ALL (21,888,015)}&100\% &7.52\% &4.85\% &87.63\% &6.14\% &0.94\% &92.92\% &0.005\% \\ \cline{2-11}
~&443&HTTPS&79.87\% &8.09\% &5.56\% &86.35\% &6.05\% &0.9\% &93.04\% &0.003\% \\ \cline{2-11}
~&80&HTTP&6.66\% &4.58\% &1.87\% &93.55\% &5.66\% &0.4\% &93.94\% &0.002\% \\ \cline{2-11}
~&993&IMAP-SSL&0.84\% &4.88\% &19.89\% &75.23\% &8.8\% &0.41\% &90.79\% &0\% \\ \cline{2-11}
~&25&SMTP&2.13\% &5.91\% &0.32\% &93.77\% &3.04\% &2.45\% &94.35\% &0.163\% \\ \cline{2-11}
~&22&SSH&1.33\% &0.28\% &0.72\% &99.00\% &6.61\% &2.71\% &90.68\% &0\% \\ \cline{2-11}
~&5223&APNs&0.09\% &48.17\% &0.47\% &51.36\% &21.44\% &0\% &78.56\% &0\% \\ \cline{2-11}
~&8443&HTTPS&0.47\% &18.48\% &0.2\% &81.32\% &6.76\% &0\% &93.24\% &0\% \\ \cline{2-11}
~&53&DNS&0.56\% &0.01\% &0.12\% &99.87\% &0\% &0\% &100\% &0\% \\ \cline{2-11}
~&5228&Android&0.09\% &0.08\% &12\% &87.92\% &5.6\% &0\% &94.4\% &0\% \\ \cline{2-11}
~&587&SMTP&0.62\% &55.79\% &1.73\% &42.48\% &6.2\% &5.6\% &88.2\% &0\% \\ \cline{2-11}
\hline  \hline  
\multirow{6}{*}{UDP}&\multicolumn{2}{c|}{ALL (81,076,465)}&100\% &- &- &- &99.92\% &0.005\% &0.07\% &0.005\% \\ \cline{2-11}
~&53&DNS&59.95\% &- &- &- &99.99\% &0.0001\% &0.007\% &0\% \\ \cline{2-11}
~&443&QUIC&15.75\% &- &- &- &99.72\% &0.001\% &0.275\% &0\% \\ \cline{2-11}
~&123&DNS&8.53\% &- &- &- &100\% &0.001\% &0\% &0\% \\ \cline{2-11}
~&6881&Game&0.69\% &- &- &- &99.99\% &0.006\% &0.002\% &0.004\% \\ \cline{2-11}
~&51413&P2P&0.54\% &- &- &- &100\% &0.0004\% &0\% &0\% \\ 
\hline  
\end{tabular}
\end{table*}  

\subsubsection{Methodology.}
We first classify ECN-negotiation succeeded, failed, and no attempt for each TCP session where we further check the ECN field in the IP headers. 
For UDP, we check the usage of the ECN field in the IP header since UDP does not have any standard for ECN negotiation.
Table~\ref{result:passive_results} shows overall statistics of traffic traces.  
We observe 227,538,956 TCP and 81,076,465 UDP flows. We list the top 10 TCP applications and top 5 port UDP applications from these flows. 
We map each port number into the corresponding protocol and the usage percentage, as shown in Table~\ref{result:passive_results}. We also show the percentage of ECN negotiation failed and succeeded rate in TCP and ECN codepoint for TCP and UDP.

\subsubsection{ECN usage in TCP and UDP}
We observed that 7.52\% of TCP flows negotiated ECN between source and destination, 4.85\% of flows failed ECN negotiation during the TCP connection establishment, and the rest (87.6\%) did not attempt ECN negotiation.
Among the TCP flows of successful ECN negotiation (7.52\%), ECT(0), ECT(1), and CE account for 92.9\%, 0.94\%, and 0.005\%, respectively.
Overall, more than 87\% of TCP flows did not attempt ECN negotiation, which is perplexing considering that most OS versions now support ECN by default, supported by our active measurement (86.4\% of ECN-enabled Web servers). We found that although most servers enable ECN for incoming packets, ECN is not requested for outgoing packets by senders by default. Another interesting observation is that Apple's APNs (Port 5223) and the default port for SMTP submission on the modern web (Port 587) heavily use ECN for their services, while more than 20\% of APNs flows are ECN bleached in IP headers. 

From the 81,076,465 UDP flows, we found some portions of flows are using ECN in ports 53 and 443. QUIC uses UDP port 443, and DNS uses UDP port 53. QUIC published as RFC 9001~\cite{rfc9000} is a UDP-based, stream-multiplexing, and encrypted transport protocol. While a recent measurement study on QUIC~\cite{zirngibl2021s} did not mention the use of ECN for QUIC, our results confirm that some QUIC flows are now using ECN. 

\section{Discussion}
Although our results discover the current ECN deployment status and violations on the path, there are still some avenues for future measurement and improvements. 

\vspace{1mm}
\noindent{\textbf{Possible causes of ECN bleaching.}}
We have revealed that there are paths to mangling ECN, and the bleaching percentages vary along with the providers. Aside from two exceptional cases of bleaching 100\%, the ratio of the bleaching ECN field varies from 3.1\% to 51.35\%. We discuss two possible explanations for this observation.

There could be a bug in the router software leading to ECN bleaching. 
We got this clue from the correlation between ECN and DSCP in their bleaching. When ECN bleaching happens, a router overwrites two bits of the ECN field and the preceding six bits of the DSCP field. 
These two fields used a single field named ToS, which is deprecated. 
To figure out the cause, we provided the list of bleaching IPs to the ISPs who own those IPs. One of the ISPs shared information about a bug in an NPU (Neural Processing Unit) that bleaches ECN bits when it rewrites DSCP.

We also conjecture that the complete bleaching is caused by proxies/middleboxes in the provider network. 
As mentioned in Section~\ref{results_1:method}, we found two providers which completely bleach ECN. One is a cellular provider, and the other is a Cloud provider. According to recent papers, the cellular network uses a performance-enhancing proxy to improve its network performance. 
Furthermore, the cloud provider operates its proxy or middlebox, balancing the load from/to the data center. 

\vspace{1mm}
\noindent{\textbf{Potential impact of ECN bleaching to L4S performances.}}
Our measurements have identified multiple ECN bleaching cases.
As mentioned in Section~\ref{subsec:lld-background}, L4S is using ECN as a classifier between a classic queue and an L4S queue.
In typical cases, if a packet has ECT(1) or CE, a scheduler puts packets into the low latency queue.
However, if an intermediate router has bleached packets marked with ECT(1), those packets will be put into the normal queue instead of the low latency queue. 
Furthermore, if CE marking is zeroed from one router, the congestion signal cannot reach the destination, potentially causing additional latency for the sender to get the loss signal. 

\vspace{1mm}
\noindent{\textbf{ECN in cellular networks.}}
Recently, ECN-based congestion control algorithms such as ABC~\cite{goyal2020abc} and ECLAT~\cite{kim2021eclat} are introduced to fix a large queue problem in cellular networks. ABC modifies a typical ECN mechanism to use two-bit feedback as a congestion signal for increasing or decreasing the congestion window (CWND). This approach allows the router to explicitly inform how many CWNDs it needs to lower or raise, rather than implicitly notifying the network congestion in the typical ECN mechanism. As a result, performance degradation such as throughput loss and large queuing caused by wide CWND fluctuations is alleviated. ECLAT strictly limits queuing delays within an acceptable time by proactively forwarding ECN feedback to ensure that the CWND does not exceed a certain level for delay guarantees. To do this, ECLAT analyzes ECN policies that determine when to transmit ECN feedback based on CWND growth pattern analysis and network calculus. As long as ECN is well supported without bleaching, these ECN-based techniques have the potential to improve the network performances in cellular networks significantly.

\section{Related Work}\label{sec:related}

\begin{table*}[t]
\renewcommand{\arraystretch}{1.2}
\caption{Comparison with prior ECN measurement studies \textmd{({\footnotesize\halfcirc} : partially meet the criterion; {\footnotesize\fullcirc} : has good results with details; Blank: not mentioned in the literature)}.}\label{tbl:ecn-comparison}
\vspace{-0.1in}
\begin{center}
\scalebox{0.95}{%
\begin{tabular}{| c | c | c | c | c | c | c | c | c |c |}
\hline
\textbf{Methodology} & 2000 & 2004 & 2011 & 2013 & 2015 & 2017 & 2018 & 2019 & Our \\
& \cite{padhye2001identifying} & \cite{craven2014middlebox}   & \cite{bauer2011measuring} & \cite{kuhlewind2013state} & \cite{mcquistin2015explicit} & \cite{learmonth2017path} & \cite{kuhlewind2018tracing} & \cite{roddav2019usage} & work \\
\hline
\hline
\textbf{Percentage of ECN-capable web servers (TCP)}      & 1.3\% & 2.2\% & 17.2\% & 29.5\% & 56.4\%& &78.5\%& & 86.4\% \\
\hline
\textbf{Percentage of ECN marking changes}      & & & 6.3\%  & 10.9\% &2.1\%& & 4.7\%& & 10.8\% \\
\hline
\textbf{Bleaching localization}& & & \halfcirc & & & & & & \fullcirc\\
\hline
\textbf{End-to-end testing}   & & & & & & & & & \fullcirc \\
\hline
\textbf{Aggregated flow data} & & & & & \halfcirc & & & \halfcirc & \fullcirc \\
\hline
\textbf{Cellular networks}    & & & & & & \halfcirc &  & & \fullcirc\\
\hline
\end{tabular}}
\end{center}
\end{table*}

\subsection{ECN Measurement Study}

Prior measurement studies, as compared in Table \ref{tbl:ecn-comparison}, have investigated the server-side ECN capability among the major web servers as ranked by Alexa and its usability (or traversal) along the paths to the servers~\cite{bauer2011measuring,kuhlewind2013state,trammell2015observing,learmonth2016pathspider,mandalari2018measuring}. Padhye \textit{et al.}~\cite{padhye2001identifying} first revealed very little (server-side) deployment ratio of ECN, which is only 1.3\% of the probed web servers. Bauer \textit{et al.}~\cite{bauer2011measuring} later presented a comprehensive measurement study of the ECN readiness in the Internet. They found that ECN was enabled in 14\%--17\% of web servers and 0.1\%--4\% of clients, and 6\%--28.4\% of paths cleared the ECN field of packets. K\"{u}hlewind \textit{et al.}~\cite{kuhlewind2013state} also tested 22,487 hosts and reported that ECN was not usable for 9\% of the hosts, due to the middleboxes along the paths to the hosts. 

In a similar vein, Trammell \textit{et al.}~\cite{trammell2015observing} found that when testing 326,743 hosts that can negotiate the ECN capability during connection establishment, 0.03\% of the hosts (i.e., 107 hosts) experienced the failure of ECN negotiation due to the paths to the hosts, which mangle the ECE and CWR flags in the TCP header. In addition, Learmonth \textit{et al.}~\cite{learmonth2016pathspider} developed a measurement tool named PATHspider to measure the Internet path transparency for various protocol features of TCP (see Section \ref{sec:method1} for more details). McQuistin \textit{et al.}~\cite{mcquistin2015explicit} conducted an Internet path transparency measurement study for UDP traffic. Learmonth \textit{et al.}~\cite{learmonth2017path} also presented a measurement study of the ECN capability on mobile access networks, but this work is limited to checking the ECN negotiation with public websites. It is worth noting that the ECN negotiation is possible, even if the paths to the websites bleach the ECN field of IP packets, as the ECN negotiation is done based on the ECE and CWR flags of the TCP header during the TCP connection establishment.

\subsection{Inferring Proxies in Cellular Networks}

There are a few measurement studies on inferring middeboxes in the Internet, especially cellular networks. Wang \textit{et al.}~\cite{wang2011untold} conducted a measurement study for commercial cellular networks and observed that various types of packet modifications happen due to middleboxes. Detal \textit{et al.}~\cite{detal2013revealing} revealed the presence of middleboxes along the Internet paths from an experimental study using Tracebox. Xu \textit{et al.}~\cite{xu2015investigating} used an experimental testbed to investigate transparent web proxies in the four major US mobile providers and see how they behave in the presence of real web workloads. Chung \textit{et al.}~\cite{chung2016tunneling} presented Luminati, an HTTP proxy network, for measuring the network infrastructure to demystify the end-to-end violations. Zullo \textit{et al.}~\cite{zullo2017hic} proposed Mobile Tracebox, which sends crafted packets to validate intermediate boxes to see if they modify the packets or alter the path between source and destination. However, none of the prior studies have evaluated the ECN violations in wired and cellular networks.

\section{Conclusion}\label{sec:conc}
In this work, we conducted a large-scale measurement study on the traversal of ECN in the Internet, involving a wide range of cellular and wired networks. In particular, we were able to provide detailed analysis results on the instances of ECN bleaching, including how often and where in the network the bleaching instances happen. We observed that 1,112 out of 129,252 routers and $4.17\%$ of paths showed ECN bleaching, and only few networks bleach as a matter of policy. We have already shared our measurement results with ISPs and received a response from some of them, showing their willingness to fix the issue with ECN bleaching.

\bibliographystyle{ACM-Reference-Format}
\bibliography{main}


\begin{thebibliography}{40}


\ifx \showCODEN    \undefined \def \showCODEN     #1{\unskip}     \fi
\ifx \showDOI      \undefined \def \showDOI       #1{#1}\fi
\ifx \showISBNx    \undefined \def \showISBNx     #1{\unskip}     \fi
\ifx \showISBNxiii \undefined \def \showISBNxiii  #1{\unskip}     \fi
\ifx \showISSN     \undefined \def \showISSN      #1{\unskip}     \fi
\ifx \showLCCN     \undefined \def \showLCCN      #1{\unskip}     \fi
\ifx \shownote     \undefined \def \shownote      #1{#1}          \fi
\ifx \showarticletitle \undefined \def \showarticletitle #1{#1}   \fi
\ifx \showURL      \undefined \def \showURL       {\relax}        \fi
\providecommand\bibfield[2]{#2}
\providecommand\bibinfo[2]{#2}
\providecommand\natexlab[1]{#1}
\providecommand\showeprint[2][]{arXiv:#2}

\bibitem[\protect\citeauthoryear{Alexa}{Alexa}{2021}]%
        {alexa}
\bibfield{author}{\bibinfo{person}{Alexa}.} \bibinfo{year}{2021}\natexlab{}.
\newblock \bibinfo{title}{Alexa Top Sites}.
\newblock   (\bibinfo{year}{2021}).
\newblock
\showURL{%
\url{https://ats.alexa.com/}}


\bibitem[\protect\citeauthoryear{Alizadeh, Greenberg, Maltz, Padhye, Patel,
  Prabhakar, Sengupta, and Sridharan}{Alizadeh et~al\mbox{.}}{2010}]%
        {10.1145/1851182.1851192}
\bibfield{author}{\bibinfo{person}{Mohammad Alizadeh}, \bibinfo{person}{Albert
  Greenberg}, \bibinfo{person}{David~A. Maltz}, \bibinfo{person}{Jitendra
  Padhye}, \bibinfo{person}{Parveen Patel}, \bibinfo{person}{Balaji Prabhakar},
  \bibinfo{person}{Sudipta Sengupta}, {and} \bibinfo{person}{Murari
  Sridharan}.} \bibinfo{year}{2010}\natexlab{}.
\newblock \showarticletitle{Data Center TCP (DCTCP)}. In
  \bibinfo{booktitle}{{\em Proceedings of the ACM SIGCOMM 2010 Conference}}
  {\em (\bibinfo{series}{SIGCOMM '10})}. \bibinfo{publisher}{Association for
  Computing Machinery}, \bibinfo{address}{New York, NY, USA},
  \bibinfo{pages}{63–74}.
\newblock
\showISBNx{9781450302012}
\showDOI{%
\url{https://doi.org/10.1145/1851182.1851192}}


\bibitem[\protect\citeauthoryear{Bauer, Beverly, and Berger}{Bauer
  et~al\mbox{.}}{2011}]%
        {bauer2011measuring}
\bibfield{author}{\bibinfo{person}{Steven Bauer}, \bibinfo{person}{Robert
  Beverly}, {and} \bibinfo{person}{Arthur Berger}.}
  \bibinfo{year}{2011}\natexlab{}.
\newblock \showarticletitle{Measuring the state of ECN readiness in servers,
  clients, and routers}. In \bibinfo{booktitle}{{\em Proceedings of the 2011
  ACM SIGCOMM conference on Internet measurement conference}}.
  \bibinfo{pages}{171--180}.
\newblock


\bibitem[\protect\citeauthoryear{Bensley, Thaler, Balasubramanian, Eggert, and
  Judd}{Bensley et~al\mbox{.}}{2017}]%
        {DBLP:journals/rfc/rfc8257}
\bibfield{author}{\bibinfo{person}{Stephen~E. Bensley}, \bibinfo{person}{David
  Thaler}, \bibinfo{person}{Praveen Balasubramanian}, \bibinfo{person}{Lars
  Eggert}, {and} \bibinfo{person}{Glenn Judd}.}
  \bibinfo{year}{2017}\natexlab{}.
\newblock \showarticletitle{Data Center {TCP} {(DCTCP):} {TCP} Congestion
  Control for Data Centers}.
\newblock \bibinfo{journal}{{\em {RFC}\/}}  \bibinfo{volume}{8257}
  (\bibinfo{year}{2017}), \bibinfo{pages}{1--17}.
\newblock
\showDOI{%
\url{https://doi.org/10.17487/RFC8257}}


\bibitem[\protect\citeauthoryear{Bhooma}{Bhooma}{2017}]%
        {Padma2027apple}
\bibfield{author}{\bibinfo{person}{Padma Bhooma}.}
  \bibinfo{year}{2017}\natexlab{}.
\newblock \bibinfo{title}{Experience with enabling ECN on the Internet,
  (https://www.ietf.org/proceedings/98/slides/slides-98-maprg-tcp-ecn-experience-with-enabling-ecn-on-the-internet-padma-bhooma-00)}.
\newblock   (\bibinfo{year}{2017}).
\newblock
\showURL{%
\url{https://tiny.one/4ea6rfvv}}


\bibitem[\protect\citeauthoryear{Briscoe, Schepper, Bagnulo, and White}{Briscoe
  et~al\mbox{.}}{2022}]%
        {ietf-tsvwg-l4s-arch-17}
\bibfield{author}{\bibinfo{person}{Bob Briscoe}, \bibinfo{person}{Koen~De
  Schepper}, \bibinfo{person}{Marcelo Bagnulo}, {and} \bibinfo{person}{Greg
  White}.} \bibinfo{year}{2022}\natexlab{}.
\newblock \bibinfo{booktitle}{{\em {Low Latency, Low Loss, Scalable Throughput
  (L4S) Internet Service: Architecture}}}.
\newblock \bibinfo{type}{Internet-Draft} draft-ietf-tsvwg-l4s-arch-17.
  \bibinfo{institution}{Internet Engineering Task Force}.
\newblock
\showURL{%
\url{https://datatracker.ietf.org/doc/html/draft-ietf-tsvwg-l4s-arch-17}}
\newblock
\shownote{Work in Progress.}


\bibitem[\protect\citeauthoryear{Brunello, Johansson, Ozger, and
  Cavdar}{Brunello et~al\mbox{.}}{2021}]%
        {brunello2021low}
\bibfield{author}{\bibinfo{person}{Davide Brunello}, \bibinfo{person}{Ingemar
  Johansson}, \bibinfo{person}{Mustafa Ozger}, {and} \bibinfo{person}{Cicek
  Cavdar}.} \bibinfo{year}{2021}\natexlab{}.
\newblock \showarticletitle{Low Latency Low Loss Scalable Throughput in 5G
  Networks}. In \bibinfo{booktitle}{{\em 2021 IEEE 93rd Vehicular Technology
  Conference (VTC2021-Spring)}}. IEEE, \bibinfo{pages}{1--7}.
\newblock


\bibitem[\protect\citeauthoryear{Chung, Choffnes, and Mislove}{Chung
  et~al\mbox{.}}{2016}]%
        {chung2016tunneling}
\bibfield{author}{\bibinfo{person}{Taejoong Chung}, \bibinfo{person}{David
  Choffnes}, {and} \bibinfo{person}{Alan Mislove}.}
  \bibinfo{year}{2016}\natexlab{}.
\newblock \showarticletitle{Tunneling for transparency: A large-scale analysis
  of end-to-end violations in the Internet}. In \bibinfo{booktitle}{{\em
  Proceedings of the 2016 Internet Measurement Conference}}.
  \bibinfo{pages}{199--213}.
\newblock


\bibitem[\protect\citeauthoryear{Craven, Beverly, and Allman}{Craven
  et~al\mbox{.}}{2014}]%
        {craven2014middlebox}
\bibfield{author}{\bibinfo{person}{Ryan Craven}, \bibinfo{person}{Robert
  Beverly}, {and} \bibinfo{person}{Mark Allman}.}
  \bibinfo{year}{2014}\natexlab{}.
\newblock \showarticletitle{A middlebox-cooperative TCP for a non end-to-end
  Internet}.
\newblock \bibinfo{journal}{{\em ACM SIGCOMM Computer Communication Review\/}}
  \bibinfo{volume}{44}, \bibinfo{number}{4} (\bibinfo{year}{2014}),
  \bibinfo{pages}{151--162}.
\newblock


\bibitem[\protect\citeauthoryear{Detal, Hesmans, Bonaventure, Vanaubel, and
  Donnet}{Detal et~al\mbox{.}}{2013}]%
        {detal2013revealing}
\bibfield{author}{\bibinfo{person}{Gregory Detal}, \bibinfo{person}{Benjamin
  Hesmans}, \bibinfo{person}{Olivier Bonaventure}, \bibinfo{person}{Yves
  Vanaubel}, {and} \bibinfo{person}{Benoit Donnet}.}
  \bibinfo{year}{2013}\natexlab{}.
\newblock \showarticletitle{Revealing middlebox interference with tracebox}. In
  \bibinfo{booktitle}{{\em Proceedings of the 2013 conference on Internet
  measurement conference}}. \bibinfo{pages}{1--8}.
\newblock


\bibitem[\protect\citeauthoryear{Duplyakin, Ricci, Maricq, Wong, Duerig, Eide,
  Stoller, Hibler, Johnson, Webb, Akella, Wang, Ricart, Landweber, Elliott,
  Zink, Cecchet, Kar, and Mishra}{Duplyakin et~al\mbox{.}}{2019}]%
        {cloudlab}
\bibfield{author}{\bibinfo{person}{Dmitry Duplyakin}, \bibinfo{person}{Robert
  Ricci}, \bibinfo{person}{Aleksander Maricq}, \bibinfo{person}{Gary Wong},
  \bibinfo{person}{Jonathon Duerig}, \bibinfo{person}{Eric Eide},
  \bibinfo{person}{Leigh Stoller}, \bibinfo{person}{Mike Hibler},
  \bibinfo{person}{David Johnson}, \bibinfo{person}{Kirk Webb},
  \bibinfo{person}{Aditya Akella}, \bibinfo{person}{Kuangching Wang},
  \bibinfo{person}{Glenn Ricart}, \bibinfo{person}{Larry Landweber},
  \bibinfo{person}{Chip Elliott}, \bibinfo{person}{Michael Zink},
  \bibinfo{person}{Emmanuel Cecchet}, \bibinfo{person}{Snigdhaswin Kar}, {and}
  \bibinfo{person}{Prabodh Mishra}.} \bibinfo{year}{2019}\natexlab{}.
\newblock \showarticletitle{The Design and Operation of {CloudLab}}. In
  \bibinfo{booktitle}{{\em Proceedings of USENIX ATC}}.
\newblock


\bibitem[\protect\citeauthoryear{Fairhurst and Welzl}{Fairhurst and
  Welzl}{2017}]%
        {rfc8087}
\bibfield{author}{\bibinfo{person}{G. Fairhurst} {and} \bibinfo{person}{M.
  Welzl}.} \bibinfo{year}{2017}\natexlab{}.
\newblock \bibinfo{title}{RFC8087: The Benefits of Using Explicit Congestion
  Notification (ECN)}.
\newblock   (\bibinfo{year}{2017}).
\newblock


\bibitem[\protect\citeauthoryear{Goyal, Agarwal, Netravali, Alizadeh, and
  Balakrishnan}{Goyal et~al\mbox{.}}{2020}]%
        {goyal2020abc}
\bibfield{author}{\bibinfo{person}{Prateesh Goyal}, \bibinfo{person}{Anup
  Agarwal}, \bibinfo{person}{Ravi Netravali}, \bibinfo{person}{Mohammad
  Alizadeh}, {and} \bibinfo{person}{Hari Balakrishnan}.}
  \bibinfo{year}{2020}\natexlab{}.
\newblock \showarticletitle{$\{$ABC$\}$: A Simple Explicit Congestion
  Controller for Wireless Networks}. In \bibinfo{booktitle}{{\em 17th USENIX
  Symposium on Networked Systems Design and Implementation (NSDI 20)}}.
  \bibinfo{pages}{353--372}.
\newblock


\bibitem[\protect\citeauthoryear{Holland}{Holland}{2020}]%
        {Holland2020}
\bibfield{author}{\bibinfo{person}{Jake Holland}.}
  \bibinfo{year}{2020}\natexlab{}.
\newblock \bibinfo{title}{Latency \& AQM Observations on the Internet,
  (https://datatracker.ietf.org/meeting/interim-2020-maprg-01/materials/slides-interim-2020-maprg-01-sessa-latency-aqm-observations-on-the-internet-01)}.
\newblock   (\bibinfo{year}{2020}).
\newblock
\showURL{%
\url{https://tiny.one/2p8pabpt}}


\bibitem[\protect\citeauthoryear{Høiland-Jørgensen}{Høiland-Jørgensen}{2016}]%
        {enable-ecn}
\bibfield{author}{\bibinfo{person}{Toke Høiland-Jørgensen}.}
  \bibinfo{year}{2016}\natexlab{}.
\newblock \bibinfo{title}{Bufferbloat.net -- Enable ECN}.
\newblock   (\bibinfo{year}{2016}).
\newblock
\newblock
\shownote{\url{https://www.bufferbloat.net/projects/cerowrt/wiki/Enable_ECN/}.}


\bibitem[\protect\citeauthoryear{Iyengar and Thomson}{Iyengar and
  Thomson}{2021}]%
        {rfc9000}
\bibfield{author}{\bibinfo{person}{Jana Iyengar} {and} \bibinfo{person}{Martin
  Thomson}.} \bibinfo{year}{2021}\natexlab{}.
\newblock \bibinfo{title}{{QUIC: A UDP-Based Multiplexed and Secure
  Transport}}.
\newblock \bibinfo{howpublished}{RFC 9000}.   (\bibinfo{date}{May}
  \bibinfo{year}{2021}).
\newblock
\showDOI{%
\url{https://doi.org/10.17487/RFC9000}}


\bibitem[\protect\citeauthoryear{Kim, Im, and Lee}{Kim et~al\mbox{.}}{2021}]%
        {kim2021eclat}
\bibfield{author}{\bibinfo{person}{Junseon Kim}, \bibinfo{person}{Youngbin Im},
  {and} \bibinfo{person}{Kyunghan Lee}.} \bibinfo{year}{2021}\natexlab{}.
\newblock \showarticletitle{ECLAT: An ECN Marking System for Latency Guarantee
  in Cellular Networks}. In \bibinfo{booktitle}{{\em IEEE INFOCOM 2021-IEEE
  Conference on Computer Communications}}. IEEE, \bibinfo{pages}{1--10}.
\newblock


\bibitem[\protect\citeauthoryear{K{\"u}hlewind, Neuner, and
  Trammell}{K{\"u}hlewind et~al\mbox{.}}{2013}]%
        {kuhlewind2013state}
\bibfield{author}{\bibinfo{person}{Mirja K{\"u}hlewind},
  \bibinfo{person}{Sebastian Neuner}, {and} \bibinfo{person}{Brian Trammell}.}
  \bibinfo{year}{2013}\natexlab{}.
\newblock \showarticletitle{On the State of ECN and TCP Options on the
  Internet}. In \bibinfo{booktitle}{{\em International Conference on Passive
  and Active Network Measurement}}. Springer, \bibinfo{pages}{135--144}.
\newblock


\bibitem[\protect\citeauthoryear{K{\"u}hlewind, Walter, Learmonth, and
  Trammell}{K{\"u}hlewind et~al\mbox{.}}{2018}]%
        {kuhlewind2018tracing}
\bibfield{author}{\bibinfo{person}{Mirja K{\"u}hlewind},
  \bibinfo{person}{Michael Walter}, \bibinfo{person}{Iain~R Learmonth}, {and}
  \bibinfo{person}{Brian Trammell}.} \bibinfo{year}{2018}\natexlab{}.
\newblock \showarticletitle{Tracing internet path transparency}. In
  \bibinfo{booktitle}{{\em 2018 Network Traffic Measurement and Analysis
  Conference (TMA)}}. IEEE, \bibinfo{pages}{1--7}.
\newblock


\bibitem[\protect\citeauthoryear{Langley}{Langley}{2008}]%
        {langley2008probing}
\bibfield{author}{\bibinfo{person}{Adam Langley}.}
  \bibinfo{year}{2008}\natexlab{}.
\newblock \showarticletitle{Probing the viability of TCP extensions}.
\newblock \bibinfo{journal}{{\em Google, Inc., Tech. Rep\/}}
  (\bibinfo{year}{2008}).
\newblock


\bibitem[\protect\citeauthoryear{Learmonth, Lutu, Fairhurst, Ros, and
  Alay}{Learmonth et~al\mbox{.}}{2017}]%
        {learmonth2017path}
\bibfield{author}{\bibinfo{person}{Iain~R Learmonth}, \bibinfo{person}{Andra
  Lutu}, \bibinfo{person}{Gorry Fairhurst}, \bibinfo{person}{David Ros}, {and}
  \bibinfo{person}{{\"O}zg{\"u} Alay}.} \bibinfo{year}{2017}\natexlab{}.
\newblock \showarticletitle{Path transparency measurements from the mobile edge
  with PATHspider}. In \bibinfo{booktitle}{{\em 2017 Network Traffic
  Measurement and Analysis Conference (TMA)}}. IEEE, \bibinfo{pages}{1--6}.
\newblock


\bibitem[\protect\citeauthoryear{Learmonth, Trammell, Kuhlewind, and
  Fairhurst}{Learmonth et~al\mbox{.}}{2016}]%
        {learmonth2016pathspider}
\bibfield{author}{\bibinfo{person}{Iain~R Learmonth}, \bibinfo{person}{Brian
  Trammell}, \bibinfo{person}{Mirja Kuhlewind}, {and} \bibinfo{person}{Gorry
  Fairhurst}.} \bibinfo{year}{2016}\natexlab{}.
\newblock \showarticletitle{PATHspider: A tool for active measurement of path
  transparency}. In \bibinfo{booktitle}{{\em Proceedings of the 2016 Applied
  Networking Research Workshop}}. \bibinfo{pages}{62--64}.
\newblock


\bibitem[\protect\citeauthoryear{Mandalari, Lutu, Briscoe, Bagnulo, and
  Alay}{Mandalari et~al\mbox{.}}{2018}]%
        {mandalari2018measuring}
\bibfield{author}{\bibinfo{person}{Anna~Maria Mandalari},
  \bibinfo{person}{Andra Lutu}, \bibinfo{person}{Bob Briscoe},
  \bibinfo{person}{Marcelo Bagnulo}, {and} \bibinfo{person}{Ozgu Alay}.}
  \bibinfo{year}{2018}\natexlab{}.
\newblock \showarticletitle{Measuring ECN++: good news for++, bad news for ECN
  over mobile}.
\newblock \bibinfo{journal}{{\em IEEE Communications Magazine\/}}
  \bibinfo{volume}{56}, \bibinfo{number}{3} (\bibinfo{year}{2018}),
  \bibinfo{pages}{180--186}.
\newblock


\bibitem[\protect\citeauthoryear{McQuistin and Perkins}{McQuistin and
  Perkins}{2015}]%
        {mcquistin2015explicit}
\bibfield{author}{\bibinfo{person}{Stephen McQuistin} {and}
  \bibinfo{person}{Colin~S Perkins}.} \bibinfo{year}{2015}\natexlab{}.
\newblock \showarticletitle{Is Explicit Congestion Notification usable with
  UDP?}. In \bibinfo{booktitle}{{\em Proceedings of the 2015 Internet
  Measurement Conference}}. \bibinfo{pages}{63--69}.
\newblock


\bibitem[\protect\citeauthoryear{Medina, Allman, and Floyd}{Medina
  et~al\mbox{.}}{2005}]%
        {medina2005measuring}
\bibfield{author}{\bibinfo{person}{Alberto Medina}, \bibinfo{person}{Mark
  Allman}, {and} \bibinfo{person}{Sally Floyd}.}
  \bibinfo{year}{2005}\natexlab{}.
\newblock \showarticletitle{Measuring the evolution of transport protocols in
  the Internet}.
\newblock \bibinfo{journal}{{\em ACM SIGCOMM Computer Communication Review\/}}
  \bibinfo{volume}{35}, \bibinfo{number}{2} (\bibinfo{year}{2005}),
  \bibinfo{pages}{37--52}.
\newblock


\bibitem[\protect\citeauthoryear{Microsoft}{Microsoft}{2021}]%
        {windows-ecn}
\bibfield{author}{\bibinfo{person}{Microsoft}.}
  \bibinfo{year}{2021}\natexlab{}.
\newblock \bibinfo{title}{Winsock explicit congestion notification (ECN)}.
\newblock   (\bibinfo{year}{2021}).
\newblock
\newblock
\shownote{\url{https://docs.microsoft.com/en-us/windows/win32/winsock/winsock-ecn}.}


\bibitem[\protect\citeauthoryear{NOKIA}{NOKIA}{2020}]%
        {wifil4s}
\bibfield{author}{\bibinfo{person}{NOKIA}.} \bibinfo{year}{2020}\natexlab{}.
\newblock \bibinfo{title}{Enjoying a real-time Internet, enabled by L4S}.
\newblock   (\bibinfo{year}{2020}).
\newblock
\showURL{%
\url{https://onestore.nokia.com/asset/207072}}


\bibitem[\protect\citeauthoryear{Padhye and Floyd}{Padhye and Floyd}{2001}]%
        {padhye2001identifying}
\bibfield{author}{\bibinfo{person}{Jitendra Padhye} {and}
  \bibinfo{person}{Sally Floyd}.} \bibinfo{year}{2001}\natexlab{}.
\newblock \bibinfo{title}{Identifying the TCP behavior of web servers}.
\newblock   (\bibinfo{year}{2001}).
\newblock


\bibitem[\protect\citeauthoryear{Ramakrishnan, Floyd, and Black}{Ramakrishnan
  et~al\mbox{.}}{2001}]%
        {ramak}
\bibfield{author}{\bibinfo{person}{K Ramakrishnan}, \bibinfo{person}{Sally
  Floyd}, {and} \bibinfo{person}{D Black}.} \bibinfo{year}{2001}\natexlab{}.
\newblock \bibinfo{title}{RFC3168: The addition of explicit congestion
  notification (ECN) to IP}.
\newblock   (\bibinfo{year}{2001}).
\newblock


\bibitem[\protect\citeauthoryear{Roddav, Streit, Rodosek, and Pras}{Roddav
  et~al\mbox{.}}{2019}]%
        {roddav2019usage}
\bibfield{author}{\bibinfo{person}{Nils Roddav}, \bibinfo{person}{Klement
  Streit}, \bibinfo{person}{Gabi~Dreo Rodosek}, {and} \bibinfo{person}{Aiko
  Pras}.} \bibinfo{year}{2019}\natexlab{}.
\newblock \showarticletitle{On the Usage of DSCP and ECN codepoints in internet
  backbone traffic traces for IPv4 and IPv6}. In \bibinfo{booktitle}{{\em 2019
  International Symposium on Networks, Computers and Communications (ISNCC)}}.
  IEEE, \bibinfo{pages}{1--6}.
\newblock


\bibitem[\protect\citeauthoryear{Schepper and Briscoe}{Schepper and
  Briscoe}{2020}]%
        {ecn2020Schepper}
\bibfield{author}{\bibinfo{person}{K. Schepper} {and} \bibinfo{person}{B.
  Briscoe}.} \bibinfo{year}{2020}\natexlab{}.
\newblock \bibinfo{title}{Identifying Modified Explicit Congestion Notification
  (ECN) Semantics for Ultra-Low Queuing Delay (L4S), Work in Progress,
  draft-ietf-tsvwg-ecn-l4s-id-12}.
\newblock   (\bibinfo{year}{2020}).
\newblock
\showURL{%
\url{http://www.ietf.org/internet-drafts/draft-ietf-tsvwg-ecn-l4s-id-12.txt}}


\bibitem[\protect\citeauthoryear{Schepper, Briscoe, and White}{Schepper
  et~al\mbox{.}}{2022}]%
        {ietf-tsvwg-aqm-dualq-coupled-23}
\bibfield{author}{\bibinfo{person}{Koen~De Schepper}, \bibinfo{person}{Bob
  Briscoe}, {and} \bibinfo{person}{Greg White}.}
  \bibinfo{year}{2022}\natexlab{}.
\newblock \bibinfo{booktitle}{{\em {DualQ Coupled AQMs for Low Latency, Low
  Loss and Scalable Throughput (L4S)}}}.
\newblock \bibinfo{type}{Internet-Draft} draft-ietf-tsvwg-aqm-dualq-coupled-23.
  \bibinfo{institution}{Internet Engineering Task Force}.
\newblock
\showURL{%
\url{https://datatracker.ietf.org/doc/html/draft-ietf-tsvwg-aqm-dualq-coupled-23}}
\newblock
\shownote{Work in Progress.}


\bibitem[\protect\citeauthoryear{Trammell and K{\"u}hlewind}{Trammell and
  K{\"u}hlewind}{2015}]%
        {trammell2015observing}
\bibfield{author}{\bibinfo{person}{Brian Trammell} {and} \bibinfo{person}{Mirja
  K{\"u}hlewind}.} \bibinfo{year}{2015}\natexlab{}.
\newblock \showarticletitle{Observing Internet path transparency to support
  protocol engineering}. In \bibinfo{booktitle}{{\em Proceedings of IRTF\/ISOC
  RAIM Workshop}}. \bibinfo{publisher}{Proceedings of IRTF\/ISOC RAIM
  Workshop}.
\newblock


\bibitem[\protect\citeauthoryear{Trammell, K{\"u}hlewind, De~Vaere, Learmonth,
  and Fairhurst}{Trammell et~al\mbox{.}}{2017}]%
        {trammell2017tracking}
\bibfield{author}{\bibinfo{person}{Brian Trammell}, \bibinfo{person}{Mirja
  K{\"u}hlewind}, \bibinfo{person}{Piet De~Vaere}, \bibinfo{person}{Iain~R
  Learmonth}, {and} \bibinfo{person}{Gorry Fairhurst}.}
  \bibinfo{year}{2017}\natexlab{}.
\newblock \showarticletitle{Tracking transport-layer evolution with
  pathspider}. In \bibinfo{booktitle}{{\em Proceedings of the Applied
  Networking Research Workshop}}. \bibinfo{pages}{20--26}.
\newblock


\bibitem[\protect\citeauthoryear{Vanaubel, Pansiot, M{\'e}rindol, and
  Donnet}{Vanaubel et~al\mbox{.}}{2013}]%
        {vanaubel2013network}
\bibfield{author}{\bibinfo{person}{Yves Vanaubel},
  \bibinfo{person}{Jean-Jacques Pansiot}, \bibinfo{person}{Pascal
  M{\'e}rindol}, {and} \bibinfo{person}{Benoit Donnet}.}
  \bibinfo{year}{2013}\natexlab{}.
\newblock \showarticletitle{Network fingerprinting: TTL-based router
  signatures}. In \bibinfo{booktitle}{{\em Proceedings of the 2013 conference
  on Internet measurement conference}}. \bibinfo{pages}{369--376}.
\newblock


\bibitem[\protect\citeauthoryear{Wang, Qian, Xu, Mao, and Zhang}{Wang
  et~al\mbox{.}}{2011}]%
        {wang2011untold}
\bibfield{author}{\bibinfo{person}{Zhaoguang Wang}, \bibinfo{person}{Zhiyun
  Qian}, \bibinfo{person}{Qiang Xu}, \bibinfo{person}{Zhuoqing Mao}, {and}
  \bibinfo{person}{Ming Zhang}.} \bibinfo{year}{2011}\natexlab{}.
\newblock \showarticletitle{An untold story of middleboxes in cellular
  networks}.
\newblock \bibinfo{journal}{{\em ACM SIGCOMM Computer Communication Review\/}}
  \bibinfo{volume}{41}, \bibinfo{number}{4} (\bibinfo{year}{2011}),
  \bibinfo{pages}{374--385}.
\newblock


\bibitem[\protect\citeauthoryear{Willars, Wittenmark, Ronkainen, Östberg,
  Johansson, Strand, Lédl, and Schnieders}{Willars et~al\mbox{.}}{2021}]%
        {5gecn2021erricsson}
\bibfield{author}{\bibinfo{person}{Per Willars}, \bibinfo{person}{Emma
  Wittenmark}, \bibinfo{person}{Henrik Ronkainen}, \bibinfo{person}{Christer
  Östberg}, \bibinfo{person}{Ingemar Johansson}, \bibinfo{person}{Johan
  Strand}, \bibinfo{person}{Petr Lédl}, {and} \bibinfo{person}{Dominik
  Schnieders}.} \bibinfo{year}{2021}\natexlab{}.
\newblock \bibinfo{title}{Enabling time-critical applications over 5G with rate
  adaptation}.
\newblock   (\bibinfo{year}{2021}).
\newblock
\showURL{%
\url{https://www.ericsson.com/en/reports-and-papers/white-papers/enabling-time-critical-applications-over-5g-with-rate-adaptation}}


\bibitem[\protect\citeauthoryear{Xu, Jiang, Flach, Katz-Bassett, Choffnes, and
  Govindan}{Xu et~al\mbox{.}}{2015}]%
        {xu2015investigating}
\bibfield{author}{\bibinfo{person}{Xing Xu}, \bibinfo{person}{Yurong Jiang},
  \bibinfo{person}{Tobias Flach}, \bibinfo{person}{Ethan Katz-Bassett},
  \bibinfo{person}{David Choffnes}, {and} \bibinfo{person}{Ramesh Govindan}.}
  \bibinfo{year}{2015}\natexlab{}.
\newblock \showarticletitle{Investigating transparent web proxies in cellular
  networks}. In \bibinfo{booktitle}{{\em International conference on passive
  and active network measurement}}. Springer, \bibinfo{pages}{262--276}.
\newblock


\bibitem[\protect\citeauthoryear{Zirngibl, Buschmann, Sattler, Jaeger, Aulbach,
  and Carle}{Zirngibl et~al\mbox{.}}{2021}]%
        {zirngibl2021s}
\bibfield{author}{\bibinfo{person}{Johannes Zirngibl},
  \bibinfo{person}{Philippe Buschmann}, \bibinfo{person}{Patrick Sattler},
  \bibinfo{person}{Benedikt Jaeger}, \bibinfo{person}{Juliane Aulbach}, {and}
  \bibinfo{person}{Georg Carle}.} \bibinfo{year}{2021}\natexlab{}.
\newblock \showarticletitle{It's over 9000: analyzing early QUIC deployments
  with the standardization on the horizon}. In \bibinfo{booktitle}{{\em
  Proceedings of the 21st ACM Internet Measurement Conference}}.
  \bibinfo{pages}{261--275}.
\newblock


\bibitem[\protect\citeauthoryear{Zullo, Pescap{\'e}, Edeline, and Donnet}{Zullo
  et~al\mbox{.}}{2017}]%
        {zullo2017hic}
\bibfield{author}{\bibinfo{person}{Raffaele Zullo}, \bibinfo{person}{Antonio
  Pescap{\'e}}, \bibinfo{person}{Korian Edeline}, {and} \bibinfo{person}{Benoit
  Donnet}.} \bibinfo{year}{2017}\natexlab{}.
\newblock \showarticletitle{Hic sunt NATs: Uncovering address translation with
  a smart traceroute}. In \bibinfo{booktitle}{{\em 2017 Network Traffic
  Measurement and Analysis Conference (TMA)}}. IEEE, \bibinfo{pages}{1--6}.
\newblock


\end{thebibliography}

\appendix
\clearpage
\section{Ethics} 
This work does not raise any ethical issues.


\end{document}